\documentclass[11pt]{article}
\usepackage{graphicx}
\usepackage{latexsym,amsmath,amsfonts,amssymb,bm,bbm}
\usepackage{jheppub}

\def\tr{{\rm tr}}

\def\CJ{{\cal J}}

\def\CL{{\cal L}}

\def\CM{{\cal M}}
\def\CN{{\cal N}}

\def\CT{{\cal T}}

\def\BH{\mathbb{H}}

\def\BP{\mathbb{P}}
\def\BR{\mathbb{R}}

\def\BZ{\mathbb{Z}}

\makeatother

\begin{document}
%
\preprint{UT-15-26}
\title{Anomalies and Entanglement Entropy}

\author[a]{Tatsuma Nishioka}
\author[b]{and Amos Yarom}

\affiliation[a]{Department of Physics, Faculty of Science,
The University of Tokyo,\\
Bunkyo-ku, Tokyo 113-0033, Japan}
\affiliation[b]{
Department of Physics, Technion, Haifa 32000, Israel}

\emailAdd{nishioka@hep-th.phys.s.u-tokyo.ac.jp}
\emailAdd{ayarom@physics.technion.ac.il}

\abstract{
We initiate a systematic study of entanglement and R{\'e}nyi entropies in the presence of gauge and gravitational anomalies in even-dimensional quantum field theories. We argue that the mixed and gravitational anomalies are sensitive to boosts and obtain a closed form expression for their behavior under such transformations. Explicit constructions exhibiting the dependence of entanglement entropy on boosts is provided for theories on spacetimes with non-trivial magnetic fluxes and (or) non-vanishing Pontryagin classes.
}

\maketitle

\section{Introduction and summary}
A remarkable feature of quantum field theory is that an admissible classical symmetry may be broken by quantum mechanical effects. When a theory possesses such a broken symmetry we refer to that symmetry as anomalous. 

There is a substantial volume of scientific works on the effect of anomalies on S-matrix elements, see, e.g., \cite{Bilal:2008qx,Harvey:2005it,Bertlmann}, for several reviews. However, little is known regarding the manifestation of anomalies in other field theoretic quantities. Recently, it has been realised that anomalies play a significant role in the thermodynamic behaviour of anomalous gauge theories \cite{Erdmenger:2008rm,Son:2009tf,Neiman:2010zi,Landsteiner:2011cp,Landsteiner:2011iq,Jensen:2012kj,Jensen:2013kka,Jensen:2013rga} leading to novel effects in heavy ion collisions (see, e.g., \cite{Kharzeev:2007jp,Kharzeev:2013ffa,Kharzeev:2015kna}), specialized Weyl semi-metals \cite{Son:2012bg,Landsteiner:2013sja,2014arXiv1412.6543L,Zhang:2015gwa} and astrophysical phenomenon \cite{Ohnishi:2014uea,Kaminski:2014jda,Shaverin:2014xya}. In this work, we focus our attention on the effect of anomalies on entanglement entropy, previously studied also in \cite{Wall:2011kb,Castro:2014tta,Guo:2015uqa,Azeyanagi:2015uoa}\,.

If the Hilbert space $\mathcal{H}$ of a quantum mechanical system can be decomposed into a direct product $\mathcal{H} = \mathcal{H}_A \otimes \mathcal{H}_{\bar{A}}$, then the entanglement entropy $S_A$ of a state $|\psi\rangle$ provides a measure of correlation between degrees of freedom of $|\psi\rangle$ in $\mathcal{H}_A$ and in $\mathcal{H}_{\bar{A}}$.
More formally, to compute the entanglement entropy one traces over the degrees of freedom in the complement $\bar A$ to obtain a density matrix $\rho_A=\tr_{\bar{A}} |\psi\rangle\langle\psi|$ in the subsystem $A$. The entanglement entropy for the region $A$ is given by the von Neumann entropy associated with $\rho_A$,
\begin{align}\label{EE_Def}
	S_A = - \tr_A \, (\rho_A \log \rho_A) \,.
\end{align}

In quantum field theory one often considers the entanglement entropy associated with a spatial region $A$. 
Starting with a state $|\psi\rangle$ the entanglement entropy is given by the von Neumann entropy of the reduced density matrix $\rho_A$ obtained by tracing over the Hilbert space of the spatial region $\bar{A}$. In what follows we will refer to the interface between $A$ and $\bar A$ as the entangling region $\Sigma = \partial A = \partial \bar{A}$ and work under the assumption that the Hilbert space of the theory can be decomposed into a tensor product of the Hilbert space associated with $A$ and that associated with $\bar A$. We will refer to (the Euclidean version of) the manifold on which the theory is placed as $\CM$.

In order to compute the entanglement entropy $S_A$ in a quantum field theory one uses \eqref{EE_Def} on a regulated version of the theory. 
When the initial state $|\psi\rangle$ is the vacuum state $|0\rangle$, the continuum limit of \eqref{EE_Def} is given by
\begin{subequations}
\label{E:replica}
\begin{equation}\label{ReplicaTrick}
	S_A =\lim_{n\to 1} S_n \,,
\end{equation}
where the R{\'e}nyi entropy, $S_n$, is given by (the analytic continuation of)
\begin{equation}
\label{E:Renyidefinition}
	S_n = - \frac{n W_1 - W_n}{n-1}\,.
\end{equation}
\end{subequations}
For integer $n$, $W_n = - \ln \int_{\CM_n} D\phi\, e^{-S}$ is the Euclidean generating function on the manifold $\CM_n$ where $\CM_n$ is the $n$-fold cover of $\mathcal{M}$ such that each sheet of the cover is connected along $A$ \cite{Calabrese:2004eu}. Therefore, $\CM_n$ has a codimension-two singularity at the entangling surface $\Sigma$ with surplus angle $2\pi (n-1)$.
For non-integer $n$, $W_n$ is defined by analytically continuing the integer valued $W_n$. Note that $\CM = \CM_1$ is the original Euclidean space without any singularity. The analytic continuation from integer $n\geq 2$ to $n=1$ used in \eqref{ReplicaTrick} is usually carried out under an implicit assumption regarding a replica symmetry, $\BZ_n$, between sheets in $\mathcal{M}_n$. See e.g., the discussions in \cite{Lewkowycz:2013nqa,Belin:2013dva,Camps:2014voa} for the role of the $\BZ_n$ symmetry in the holographic formula of entanglement entropy \cite{Ryu:2006bv,Ryu:2006ef}.
In the remainder of this work we will assume that the analytic continuation from non-integer $n$ to integer $n$ can be carried out and take \eqref{ReplicaTrick} to be the definition of the entanglement entropy in quantum field theory. 
Our conventions for Wick rotation are summarized in appendix \ref{A:conventions}.

From a formal perspective the definition \eqref{ReplicaTrick} implies that entanglement entropy is associated with the Euclidean partition function of the theory in the limit where the manifold on which the theory is defined becomes regular. It is then perhaps not surprising that the entanglement entropy is sensitive to anomalies.

Indeed, let us consider a theory with gauge, gravitational, and mixed gauge-gravitational anomalies. By gauge and gravitational anomalies we mean anomalies associated with turning on a background (non-dynamical) source term for the current and stress-energy tensor. In what follows we will refer to these source terms as a gauge-field and metric respectively, but we emphasize that these are non-dynamical fields and the associated anomalous symmetries are global. Anomalies associated with local symmetries generate non-unitary theories at best and will not be discussed further. For simplicity, we focus on Abelian gauge anomalies but our results can be easily generalized to non-Abelian ones.

In the presence of gravitational or mixed anomalies the stress-energy tensor $T^{\mu\nu}$ is no longer conserved. Instead, one finds the non-conservation law
\begin{equation}
\label{E:anomalousTmn}
	\nabla_{\mu}T^{\mu\nu} - F^{\nu\mu}J_{\mu}  = -i \tau^{\nu} \,,
\end{equation}
where $F^{\mu\nu}$ is a field strength associated with possible external fields, $J_{\mu}$ is the current associated with a global Abelian symmetry and $\tau^{\mu}$ denotes the anomalous contribution to the (non-)conservation law of the stress-energy tensor. The term $F^{\nu\mu}J_{\mu}$ is associated with Joule heating.
The explicit form of $\tau^{\mu}$ can be computed using the Wess-Zumino consistency conditions. It is given by
\begin{equation}
\label{E:deftau}
	\tau^{\mu} =-\mathcal{J}A^{\mu} - \frac{1}{\sqrt{g}}g^{\mu\nu}\partial_{\rho} \left(\sqrt{g}\, \mathcal{T}^{\rho}{}_{\nu}\right) \,,\\
\end{equation}
where $A_{\mu}$ is the gauge field associated with $F_{\mu\nu}$ and $\mathcal{J}$ and $\mathcal{T}^{\mu}{}_{\nu}$ may be expressed as variations of the Chern-Simons forms associated with the anomaly. In the presence of boundaries \eqref{E:deftau} is expected to receive corrections \cite{PSP:2075184,PSP:2075856,PSP:2128256}. We provide a derivation of \eqref{E:deftau} in section \ref{s:EEAnom}. A derivation of the explicit expressions for $\mathcal{J}$ and $\mathcal{T}^{\mu}{}_{\nu}$ can be found in appendix \ref{A:inflow} and in \cite{Jensen:2013kka}. 
It is worth emphasizing that $\tau^{\mu}$ in \eqref{E:deftau} specifies the non-conservation law for the {consistent} stress tensor in Euclidean signature. The expression for $\tau$ associated with the covariant stress tensor is somewhat different.

Consider boosting the state $|0\rangle$ by a hyperbolic angle $\kappa$ which implies a rotation of the Euclidean vacuum by an angle $\theta = i \kappa$. Let $\xi^{\mu}\partial_{\mu}$ be the generator of this rotation. We argue that if the dynamics are invariant under $\theta$ in the absence of an anomaly, then the R{\'e}nyi entropy associated with the boosted state $|0\rangle$ in the presence of the anomaly satisfies 
\begin{equation}
\label{E:Snmain}
	\partial_{\theta}S_n\big|_{\theta=0} = -\frac{i}{n-1}\left[ n \int_{\mathcal{M}_1} d^dx\, \sqrt{g}\,\partial_{\nu}\xi^{\mu}\, \mathcal{T}^{\nu}{}_{\mu} - \int_{\mathcal{M}_n} d^dx \sqrt{g}\, \partial_{\nu}\xi^{\mu}\, \mathcal{T}^{\nu}{}_{\mu}\right] \,.
\end{equation}
Equation \eqref{E:Snmain} is valid in the absence of boundaries. We discuss possible subtle corrections to \eqref{E:Snmain} for manifolds with boundaries or coordinate singularities in section \ref{S:discussion}.
A similar expression for the entanglement entropy can be obtained by inserting \eqref{E:Snmain} into \eqref{ReplicaTrick}.
In what follows we will omit the $\big|_{\theta}$ symbol, keeping in mind that all our results for derivatives of the entanglement entropy are evaluated at $\theta=0$. The imaginary contribution to the entanglement entropy follows from the analytic continuation $\kappa \to -i\theta$ and is removed when analytically continuing back to the Lorentzian signature manifold.

We carry out an explicit evaluation of \eqref{E:Snmain} in configurations where the entangling surface splits space into two regions, i.e., $A = \{\vec{x} | x^1>0\}$. In this case the (Euclidean) metric on $\CM_n$ is given by $ds^2 = d\rho^2 + \rho^2 d\tau^2+\sum_{i=2}^{d-1} g_{ij}(x)\,dx^i dx^j$ where $0 \leq \tau < 2\pi n$. Such a geometry describes a cone with opening angle $2\pi n$. Using \eqref{E:Snmain} we obtain the following results:
for a two-dimensional quantum field theory and the vacuum state we have
\begin{equation}
\label{E:2dresult}
	\partial_{\theta}S_n = 2 \pi i\, c_g \left(1+\frac{1}{n} \right) \,,
\end{equation}
leading to
\begin{equation}
\label{E:2dresult_Interval}
	\partial_{\theta}S_A = 4 \pi i \,c_g \,,
\end{equation}
where $c_g = -\sum_i \chi_i/(96\pi)$ is the strength of the anomaly such that the index $i$ runs over all chiral particle species with $\chi_i$ denoting the chirality. For a conformal field theory $c_g = (c_L-c_R)/(96\pi)$. In order to make contact with the existing literature we argue, in section \ref{SS:twod}, that if we use a finite interval instead of a semi-infinite line for the entangling surface then \eqref{E:2dresult_Interval} will be modified to $\partial_{\theta}S_A = 8\pi i\,c_g$. A comparison between our field theory results and a holographic computation can be found in appendix \ref{A:holography}.  A more elaborate discussion and comparison can be found in section \ref{S:discussion} where we discuss possible corrections due to boundary terms.

For a four-dimensional theory and in the presence of an external magnetic field, $F_{23}=B$, through the entangling surface, we obtain
\begin{equation}
\label{E:4dresult_Intro}
	\partial_{\theta}S_A = 4\pi i\,\alpha \, c_mB\,\text{Vol}_\Sigma \,,
\end{equation}
where $c_m=\sum_i \chi_i q_i/(192\pi^2)$ is the strength of the anomaly (with $q_i$ denoting the fermion charge and $\chi_i$ the chirality) and $\alpha$ denotes an ambiguity in the Chern-Simons term derived from the mixed anomaly polynomial $\bm{F} \wedge \bm{R} \wedge \bm{R}$ and $\text{Vol}_\Sigma$ is the volume of the entangling surface $\Sigma$.
If we choose the contact terms in the theory such that the anomaly is manifest only in the gravitational sector, then $\alpha=1$.

For a six-dimensional theory in Minkowski space and with a magnetic field $F_{23}= B_1$ and $F_{45}=B_2$, we obtain
\begin{equation}
\label{E:6dresultA}
	\partial_\theta S_A = 4\pi i\,\alpha\, c_m B_1 B_2 \,\text{Vol}_\Sigma \,,
\end{equation}
where $c_m= \sum_i \chi_i q_i^2 / (768\pi^3)$ is the strength of the mixed anomaly in six dimensions (with $\chi_i$ and $q_i$ denoting the chirality and charge of fermion species). If we place the theory on $\mathbb{R}^{2} \times K3$, we find
\begin{equation}
\label{E:6dresultB}
	\partial_{\theta}S_A =1536  \pi^3i\, c_a\,,
\end{equation}
where now $c_a=-\sum_i (\chi_i - 8 t_i)/(36864 \pi^3)$ (with $\chi_i$ and $t_i$ counting fermion species chirality and self-dual and anti-self-dual two-form field species) is the strength of one of the two possible gravitational anomalies of a six-dimensional theory. It seems that in order to obtain a non-zero value for the entanglement entropy of the other gravitational anomaly one needs to consider more intricate geometries. In section \ref{SS:higherd} we present results for higher-dimensional theories.

While this work was being completed we became aware of \cite{IqbalWall} where it was pointed out that coordinate singularities may become physical once gravitational anomalies are present. We thank the authors of \cite{IqbalWall} for sharing their draft with us prior to publication. We discuss the possible role of such corrections in section \ref{S:discussion}.

\section{Entanglement and anomalies}\label{s:EEAnom}

Equation \eqref{ReplicaTrick} implies that the entanglement entropy is associated with the limiting behaviour of the generating function as the manifold it is placed on becomes non-singular. Indeed, the behaviour of entanglement entropy on scaling transformations of the entangling surface can be tied to the scaling behaviour of the generating function $W_n$ which, in turn, depends on the central charges of the theory if the latter is conformal \cite{Holzhey:1994we,Ryu:2006ef,Solodukhin:2008dh,Myers:2010tj,Casini:2011kv}. In particular, if we denote a scaling variation of the entangling surface by $\delta_\sigma$ then
\begin{subequations}
\label{E:traceanomaly}
\begin{equation}
	\delta_{\sigma}S_A = -\lim_{n\to 1}\frac{n \delta_{\sigma}W_1 - \delta_{\sigma}W_n}{n-1}\,.
\end{equation}
The scaling $\delta_{\sigma}W_n$ is related to the trace of the stress tensor,
\begin{equation}
\label{E:weylanomaly}
	\delta_{\sigma}W_n = \frac{1}{2} \int d^dx \sqrt{g}\, \sigma\, T^{\mu}{}_{\mu}\,,
\end{equation}
which is related to the central charges, $c_i$ and $a$, of the theory when the latter is conformally invariant, 
\begin{equation}
	T^{\mu}{}_{\mu} = 2 (-1)^{d/2}a\, E - \sum_i c_i \,I_i \,.
\end{equation}
\end{subequations}
By evaluating the curvature terms $I_i$ and $E$ on a spherical entangling surface one can infer the dependence of the entanglement entropy on the central charges directly from \eqref{E:traceanomaly}. See \cite{Ryu:2006ef,Solodukhin:2008dh,Myers:2010tj,Fursaev:2013fta} for details.

Likewise, we can consider the behaviour of the generating function under a coordinate transformation which generates an isometry, e.g., a boost or rotation. If we denote the generator of boosts by $\delta_\theta$ then
\begin{equation}
\label{E:Snboost}
	\delta_{\theta}S_n = -\frac{n \delta_{\theta}W_1 - \delta_{\theta}W_n}{n-1}\,.
\end{equation}
The behavior of $W_n$ under boosts (and coordinate transformations in general) is fixed by the Wess-Zumino consistency condition. In appendix \ref{A:inflow} we argue that for a small coordinate transformation $x^{\mu} \to x^{\mu} + \chi^{\mu}$ one has
\begin{equation}
\label{E:Wnmain1}
	\delta_{\chi} W_n = i\int d^dx \sqrt{g}\, \partial_{\nu}\chi^{\mu}\, \mathcal{T}^{\nu}{}_{\mu}\,,
\end{equation}
with
\begin{align}
\begin{split}
\label{E:defT}
	* \bm{\mathcal{T}}^{\mu}{}_{\nu} &= \, \frac{\partial \bm{I}_\text{CS}}{\partial \bm{\Gamma}^{\nu}{}_{\mu}} \,,\\
\end{split}
\end{align}
where $\bm{I}_\text{CS}$ is the (Euclidean) Chern-Simons form associated with the anomaly, $\bm{\Gamma}^{\mu}{}_{\nu}$ is related to the Christoffel connection $\Gamma^{\mu}{}_{\nu\sigma}$ via $\bm{\Gamma}^{\mu}{}_{\nu} = \Gamma^{\mu}{}_{\nu\sigma}dx^{\sigma}$ and $*\bm{\mathcal{T}}^{\mu}{}_{\nu}$ is the volume form dual to $\mathcal{T}^{\mu}{}_{\nu}$.
The derivative $\partial/\partial \bm{\Gamma}$ is taken at constant $\bm{R}$ and picks up a minus sign when acting on $p$-forms similar to the exterior derivative $d$. 

The energy-momentum tensor and conserved current can be obtained by varying the generating function with respect to the metric and gauge field respectively. Thus, under a small coordinate transformation parameterized by $\chi^{\mu}$ we have
\begin{align}
\begin{split}
\label{E:Wdiffeo}
	\delta_{\chi} W_n = & \int d^dx \sqrt{g} \left( \frac{1}{2} T^{\mu\nu} \delta_{\chi} g_{\mu\nu} + J^{\mu} \delta_{\chi}A_{\mu} \right)\,, \\
		=& \int d^dx \sqrt{g} \left[ \nabla_{\mu}\chi_{\nu} \,T^{\mu\nu} +  J^{\mu}  \left(\chi^{\nu}F_{\nu\mu} + \partial_{\mu} (\chi^{\nu}A_{\nu}) \right)  \right]\,,\\
		=& - \int d^dx \sqrt{g} \,\chi_{\nu}\left(\nabla_{\mu} T^{\mu\nu}  - F^{\nu\mu} J_{\mu}  + A^{\nu} \nabla_{\mu}J^{\mu} \right)\,,
\end{split}
\end{align}
where in the last step we have integrated by parts. Comparing \eqref{E:Wdiffeo} to \eqref{E:Wnmain1} yields \eqref{E:deftau} where we have defined
\begin{equation}
\label{divJ}
	\nabla_{\mu}J^{\mu} =-i \mathcal{J} \,.
\end{equation}
A non-zero $\mathcal{J}$ implies that the theory is not invariant under gauge transformations. Indeed, carrying out an analysis similar to the one carried out in appendix \ref{A:inflow} but for gauge transformations results in
\begin{equation}
	*\bm{\mathcal{J}}  = \frac{\partial \bm{I}_\text{CS}}{\partial \bm{A}{}_{\mu}}\,.
\end{equation}
See, e.g., \cite{Jensen:2013kka} for a full derivation.

We note in passing that in the presence of boundaries, \eqref{E:Wnmain1} and \eqref{E:Wdiffeo} receive corrections leading to a modification of our main result. On one hand, boundary terms generated by integrating \eqref{E:Wdiffeo} by parts will contribute to the conservation law in addition to a possible conservation law for the boundary stress tensor. On the other hand, a boundary term may appear in the right hand side of \eqref{E:Wnmain1} due to the anomaly. Such a boundary term may be determined by solving the Wess-Zumino consistency condition.

In constructing  the generating function $W_n$ one must deal with ambiguities associated with possible contact terms (often referred to as Bardeen counterterms \cite{Bardeen:1984pm}) which can shift the appearance of the anomaly in non-conservation laws. For instance, in the presence of a mixed gauge-gravitational anomaly, one can choose an appropriate Bardeen counterterm so that the anomaly will manifest itself entirely in the gravitational sector. A more familiar setup where a Bardeen counterterm is present is the (mixed) axial-vector anomaly; the existence of the Bardeen counterterm is what allows us to enforce the current conservation in the vector sector. In the current context the ambiguities relevant to our computation are captured by a choice of the Chern-Simons form $\bm{I}_\text{CS}$. We will provide explicit examples of such in the following subsections. An extensive discussion of these terms in the language of the current work can be found in, for example, \cite{Jensen:2013vta}.

We now choose $\chi^{\mu}$ to specify a boost by a (small) constant hyperbolic angle $\delta\theta$ in the $\tau$ rapidity direction.
We would like to compare the entanglement and R{\'e}nyi entropies in boosted and unboosted entangling regions. That is, we would like to compute, say, the R{\'e}nyi entropy for an unboosted entangling region and compare it to the R{\'e}nyi entropy for a boosted one while keeping the cutoff parameters fixed. When constructing the $n$-fold cover of the boosted entangling surface, each sheet of the covering will be shifted by an angle $\delta\theta$ such that, overall, $\mathcal{M}_n$ rotates by an angle $n\delta\theta$ (implying that rotating by a Euclidean angle $\theta = 2\pi$ will bring us back to the original state). Thus,   
\begin{equation}
\label{E:deltatod}
	\delta_{\chi}W = W(\theta+n\delta\theta) - W(\theta) = n \delta \theta \frac{\partial W}{\partial \theta}  \,.
\end{equation}
In what follows, we will denote $\chi^{\mu}= n \delta \theta \xi^{\mu}$. Inserting \eqref{E:deltatod} into \eqref{E:Wnmain1} and using \eqref{E:Snboost} we obtain \eqref{E:Snmain}. 

The careful reader will note that in the coordinate system in which $\xi^{\mu}\partial_{\mu} = \partial_{\tau}$ one has $\partial_{\alpha}\xi^{\beta} = 0$. For this reason we will work exclusively in a Cartesian coordinate system where $\partial_{\alpha}\xi^{\beta} \neq 0$. The lack of invariance under coordinate transformations is a hallmark of gravitational anomalies. Indeed, the non-tensorial behavior of $\mathcal{T}^{\mu}{}_{\nu}$ implies that the stress tensor derived from the generating function $W_n$ (referred to as the consistent stress tensor) does not transform like a tensor under coordinate transformations. It is possible to define a covariant stress tensor by adding to the consistent stress tensor a polynomial which is local in the external fields. Such a stress tensor however can not be derived from a local generating function and therefore its role in entanglement entropy is somewhat obscure.

In the remainder of this work, we evaluate the change in the R{\'e}nyi entropies, \eqref{E:Snmain}, for geometries of the form $\mathbb{R}^{2} \times \mathcal{N}$, and the particular case of an entangling region which divides space into two equal halves, $A = \{ \vec{x} | x^1>0 \}$ where the $x^1$ direction is along the spatial part of the Lorentzian continuation of $\mathbb{R}^{2}$ into $\mathbb{R}^{1,1}$, i.e.,
\begin{equation}
\label{E:ourmanifoldC}
	ds^2 =dt^2 + (dx^1)^2 + ds^2_{\mathcal{N}}\,,
\end{equation}
where $ds^2_{\CN} = \sum_{i=2}^d g_{ij}(x)\,dx^i dx^j$ is the metric of the $(d-2)$-dimensional manifold parameterized by the coordinates $x_i$ $(i=2,3\cdots, d-1)$. We will consider slightly more complicated geometries in a future publication. The metric \eqref{E:ourmanifoldC} describes a cone of opening angle $2\pi n$ and can be rewritten in the form 
\begin{equation}
\label{E:ourmanifoldP}
	ds^2 = d\rho^2 + \rho^2 d\tau^2 +  ds^2_{\mathcal{N}} \,,
\end{equation}
where $\tau$ has periodicity $2\pi n$.
The (Euclidean) cone located at $\rho = 0$ needs to be regulated in order to carry out explicit computations. In what follows we will smoothen the tip of the cone by using a regulating function $u$ such that the metric takes the form \cite{Fursaev:1995ef}
\begin{equation}
\label{E:regulatedP}
	ds^2 = u(\rho) d\rho^2 + \rho^2 d\tau^2 +  ds^2_{\mathcal{N}} \,,
\end{equation}
where $u(0) = n^2$ and $u$ asymptotes to unity at large $\rho$ outside a small region of radius $\varepsilon$ such that $u(\rho)=1$ and $u'(\rho)=0$ for $\rho > \varepsilon$. See \cite{Fursaev:1995ef}  for details. 
Going back to Cartesian coordinates we find that the regularized cone metric takes the form
\begin{equation}
\label{E:regulatedC}
	ds^2 = \frac{(t dx^1 - x^1 dt)^2 + u(t,x^1) (t dt + x^1 dx^1)^2}{t^2 + (x^1)^2} + ds^2_{\mathcal{N}}\,.
\end{equation}
Note that the origin $t=x^1=0$ is not covered by either \eqref{E:regulatedP} or \eqref{E:regulatedC}. Had we used the coordinate system advocated by \cite{Lewkowycz:2013nqa}
\begin{equation}
\label{E:nocs}
	ds^2 = \left(t^2+ (x^1)^2 + a^2\right)^{\epsilon} \left(dt^1 + d(x^1)^2\right) \,,
\end{equation}
with  $\epsilon=n-1$ and a regularization parameter $a$, as we do in appendix \ref{A:example} then there is no coordinate singularity at the origin.

Following our conventions for Wick rotation, \eqref{E:boostconvention}, the generator of rotations $\xi^{\mu}\partial_{\mu}$ (boosts analytically continued to Euclidean signature) can be related to $\chi^{\mu}\partial_{\mu}$ via
\begin{equation}
	\chi^{\mu}\partial_{\mu} = -n \delta\theta \left(t \partial_x - x \partial_t\right) = n \delta\theta \,\xi^{\mu}\partial_{\mu}\,,
\end{equation}
implying
\begin{equation}
\label{E:dxi}
	\partial_{\nu}\xi^{\mu} = \begin{pmatrix} \,0\, & \,-1\, \\ \,1\, & \,0\, \end{pmatrix}\,.
\end{equation}
Inserting \eqref{E:dxi} into \eqref{E:Wnmain1} we find that
\begin{equation}
\label{E:deltaWn}
	\delta_{\theta}W_\chi = i n \delta\theta \int dt dx^1 \int d^{d-2}x \sqrt{g} \left(-\mathcal{T}^{t}{}_{x^1} + \mathcal{T}^{x^1}{}_{t}\right)\,.
\end{equation}
A further simplification of \eqref{E:Snmain} can be obtained by noting that on $\mathcal{M}_1$ there is no conical singularity and therefore $\mathcal{T}^{\mu}{}_{\nu}$ vanishes there,
\begin{equation}
\label{E:simplification}
	\int_{\mathcal{M}_1} d^dx \sqrt{g}\, \partial_{\nu}\xi^{\mu}\, \mathcal{T}^{\nu}{}_{\mu} = 0\,.
\end{equation}
Inserting \eqref{E:deltaWn} and \eqref{E:simplification} into \eqref{E:Snmain} we find that, in the geometries we are considering,
\begin{align}
\begin{split}
\label{E:simplifiedformula}
	\partial_{\theta} S_n\big |_{\theta=0} = \frac{ i }{n-1}\int_{\mathcal{M}_n} d^dx \sqrt{g} \left(-\mathcal{T}^{t}{}_{x^1} + \mathcal{T}^{x^1}{}_{t}\right)\,.
\end{split} 
\end{align}
We now proceed to compute \eqref{E:simplifiedformula} in various dimensions.

\subsection{Two-dimensional theories}
\label{SS:twod}
For two-dimensional theories there is a single type of gravitational anomaly whose Chern-Simons term $\bm{I}_{3}$ can be derived from the four-dimensional anomaly polynomial
\begin{equation}
\label{E:I3def}
	c_g\, \bm{R}^{\mu}{}_{\nu} \wedge \bm{R}^{\nu}{}_{\mu} = c_g\,  d \left[ \mathbf{\Gamma}^\mu{}_{\nu} \wedge \bm{R}^\nu{}_\mu - \frac{1}{3} \mathbf{\Gamma}^\mu{}_\nu \wedge \mathbf{\Gamma}^\nu{}_\rho\wedge \mathbf{\Gamma}^\rho{}_\mu\right]  = d \bm{I}_3 \,,
\end{equation}
where $\bm{R}^{\mu}{}_{\nu} = \frac{1}{2} R^{\mu}{}_{\nu\rho\sigma}dx^{\rho}\wedge dx^{\sigma}$ is the Riemann tensor two-form and $\bm{\Gamma}^{\mu}{}_{\nu} = \Gamma^{\mu}{}_{\nu\rho}dx^{\rho}$ is the Christoffel one-form. The two-forms are related by $\bm{R}^{\mu}{}_{\nu} = d\bm{\Gamma}^{\mu}{}_{\nu} + \bm{\Gamma}^{\mu}{}_{\rho} \wedge \bm{\Gamma}^{\rho}{}_{\nu}$. The multiplicative constant $c_g$ can be determined explicitly from the field content of the theory,
\begin{equation}
	c_g = -\frac{1}{96 \pi} \sum_i \chi_i \,,
\end{equation}
where the sum is over all species and $\chi_i$ is the chirality of the fermions and scalars of the two-dimensional theory (Majorana-Weyl fermions contribute $\pm1/2$). For a conformal field theory one has
\begin{equation}
	c_g = \frac{c_L-c_R}{96\pi}\,,
\end{equation}
where $c_L$ and $c_R$ are the left and right central charges respectively.
In what follows we find it convenient to omit the wedge product and replace the spacetime indices with a trace so that \eqref{E:I3def} takes the form
\begin{equation}
    c_g\, \hbox{Tr}\left( \bm{R}^2 \right) = c_g\, d\,\hbox{Tr} \left(\bm{\Gamma}\bm{R} - \frac{1}{3} \bm{\Gamma}^3 \right) = d \bm{I}_3 \,.
\end{equation}

The explicit form of $\mathcal{T}^{\mu}{}_{\nu}$ can be obtained from \eqref{E:defT} as
\begin{equation}
	*\bm{\mathcal{T}}^{\mu}{}_{\nu} =c_g\, \frac{\partial}{\partial \bm{\Gamma}} \hbox{Tr} \left( \bm{\Gamma} \bm{R} - \frac{1}{3} \bm{\Gamma}^3 \right) = c_g\, d \bm{\Gamma} \,,
\end{equation}
resulting in
\begin{align}\label{E:Tau2d}
	\CT^\mu{}_\nu = c_g \epsilon^{\alpha\beta} \partial_\alpha \Gamma^\mu{}_{\nu\beta} \,. 
\end{align}
Evaluating \eqref{E:dxi} and \eqref{E:Tau2d} on the regularized cone \eqref{E:regulatedP} one finds
\begin{align}
\begin{split}
\label{E:twodresult}
	\int \partial_{\mu}\xi^{\nu} \sqrt{g}\, \mathcal{T}^{\mu}{}_{\nu} 	dt dx^1
	&= -\int \frac{c_g u'}{\rho u^2} dt dx^1 \,,
\end{split}
\end{align}
where $\rho^2 = t^2 + (x^1)^2$ and we have used $\epsilon^{tx^1} = 1/\sqrt{g}$. 
Using \eqref{E:deltaWn} this yields
\begin{align}\label{E:EE_Consistent2d}
	\delta_\theta W_n = 2\pi (n^2-1) i\delta \theta \,c_g \,.
\end{align}
The complex value of $\delta_{\theta}W_n$ is expected given that the analytic continuation of $-i\delta\theta$ back to Lorentzian signature is given by
\begin{align}
	\delta_\theta W_n \to -2\pi (n^2-1) \delta \kappa \,c_g \,.
\end{align}	
where $\delta \kappa$ is an infinitesimal boost parameter.

Inserting \eqref{E:twodresult} into \eqref{E:simplifiedformula}, we obtain 
\begin{equation}
\label{E:2dresultSnsecond}
	\partial_{\theta}S_n = 2 \pi i\, c_g \left(1+\frac{1}{n} \right)\,,
\end{equation}
which implies 
\begin{equation}
\label{E:2dresultSAvsecond}
	\partial_{\theta}S_A = 4 \pi i\, c_g \,,
\end{equation}
as argued in \eqref{E:2dresult_Interval}.

We note that it is also possible to use our formalism to compute the change in the entanglement entropy on a finite interval of length $L$. To this end, consider a manifold $\mathcal{M}_n$ with two conical singularities such that the metric takes the form
\begin{equation}
	ds^2 = dt^2 + (dx^1)^2 \,,
\end{equation}
away from $t^2 + (x^1)^2 < \varepsilon$ and away from $t^2 + (x^1 - L)^2 <\varepsilon$,
\begin{equation}
\label{E:regularizedin0}
	ds^2 = \frac{\left(t dx^1 - x^1 dt\right)^2 + u\left(\sqrt{t^2 + (x^1)^2}\right)\,\left(t dt + x^1 dx^1\right)^2  }{t^2 + (x^1)^2} \,,
\end{equation}
for $t^2 + (x^1)^2 < \varepsilon$ and
\begin{align}
\label{E:regularizedinL}
	ds^2 =\frac{\left(t dx^1 - (x^1-L) dt\right)^2 + u\left(\sqrt{t^2 + (x^1-L)^2}\right)\left(t dt + (x^1-L) dx^1\right)^2  }{t^2 + (x^1-L)^2} \,,
\end{align}
for $t^2 + (x^1-L)^2 <\varepsilon$. The line elements for \eqref{E:regularizedin0} and \eqref{E:regularizedinL} have been obtained from \eqref{E:regulatedP} with appropriate shifts.  Since $\mathcal{T}^{\mu}{}_{\nu}$ vanishes for  $t^2 + (x^1)^2 > \varepsilon$ and $t^2 + (x^1 -L)^2 > \varepsilon$ we need to evaluate \eqref{E:deltaWn} only near the conical singularities. A quick computation yields
\begin{equation}
	\partial_{\theta}S_n = 4 \pi i\, c_g \left(1+\frac{1}{n} \right)\,,
\end{equation}
which implies 
\begin{equation}
\label{E:finterval}
	\partial_{\theta}S_A = 8 \pi i\, c_g \,.
\end{equation}
In appendix \ref{A:holography} we compare the results of this section to a holographic computation.

\subsection{Four-dimensional theories}
Four-dimensional theories do not have a gravitational anomaly but may have a mixed gauge-gravitational anomaly. Thus, we may write 
\begin{equation}
\label{E:5dCS}
	c_m\,\bm{F}\, \hbox{Tr} \left( \bm{R}^2 \right) = c_m\,d \left[ \alpha \,\bm{F} \hbox{Tr} \left(\bm{\Gamma}\bm{R} - \frac{1}{3} \bm{\Gamma}^3 \right) + (1-\alpha) \bm{A}\, \hbox{Tr}\left(\bm{R}^2\right) \right] = d \bm{I}_5\,.
\end{equation}
The parameter $\alpha$ is a free parameter which manifests the possibility of adding a (Bardeen) contact term to the generating function and $c_m$ is the strength of the mixed anomaly given by
\begin{equation}
	c_m = \frac{1}{192\pi^2}\sum_i \chi_i q_i \,,
\end{equation}
with $\chi_i$ and $q_i$ the chirality and charge respectively of the fermion species in the theory.
As we will see shortly, such a contact term offers the possibility of shifting the anomaly entirely into the gauge sector of the theory.
Indeed, using \eqref{E:deftau}, one has
\begin{align}
\begin{split}
\label{4dJT}
	\CT^\mu{}_{\nu}& = \alpha \frac{c_m}{2} \epsilon^{\rho\sigma\alpha\beta} F_{\rho\sigma}\, \partial_\alpha \Gamma^\mu{}_{\nu\beta} \,,\\
    \CJ & = (1-\alpha)\frac{c_m}{4} \epsilon^{\rho\sigma\alpha\beta}R^{\nu}{}_{\lambda\rho\sigma}R^{\lambda}{}_{\nu\alpha\beta}\,,
\end{split}
\end{align}
implying a (non-)conservation law for the (Euclidean) stress tensor of the form
\begin{equation}
    \nabla_{\mu}T^{\mu\nu} = F^{\nu\mu}J_{\mu} + i \mathcal{J}A^{\mu} + \frac{i}{\sqrt{g}}g^{\mu\nu}\partial_{\rho} \left(\sqrt{g}\, \mathcal{T}^{\rho}{}_{\nu}\right)\,.
\end{equation}
The $U(1)$ current $J^{\mu}$ coupled to the external source $A^{\mu}$ satisfies the (non-)conservation law
\begin{equation}
    \nabla_{\mu}J^{\mu} = -i \mathcal{J}\,.
\end{equation}
When $\alpha=1$ the current is conserved and the anomaly is manifest only in the energy-momentum tensor.

If we are to use \eqref{E:Snmain} then we need that $\mathcal{T}^{\mu}{}_{\nu} \neq 0$. To this end  we consider the entanglement entropy associated with a state $|0 \rangle_{B}$ where $B = F_{23}$ is a magnetic field in the $x^1$ direction. In this case,
\begin{equation}
\label{E:Tau4dpartial}
    \int dx^2 dx^3 \,\mathcal{T}^{\mu}{}_{\nu} = \alpha\, c_m\Phi\, \epsilon^{23\alpha\beta} \partial_{\alpha}\Gamma^{\mu}{}_{\nu\beta} \,,
\end{equation}
with $\Phi$ the magnetic flux through $\mathcal{N}$ at $x^1=0$, i.e., the entangling surface. Denoting the volume of the entangling surface by $\text{Vol}_\Sigma$, the flux can be written in the form $\Phi = B\, \hbox{Vol}_{\Sigma}$. Expression \eqref{E:Tau4dpartial} is identical to \eqref{E:Tau2d} upon identifying $c_g = \alpha\, c_m\, \Phi$. Thus,
\begin{equation}
\label{E:4dresult}
    \partial_{\theta}S_n = 2 \pi i\, \alpha\, c_mB\, \left(1+\frac{1}{n}\right) \hbox{Vol}_{\Sigma}  \,,
\end{equation}
leading to
\begin{equation}
    \partial_{\theta}S_A = 4 \pi i\, \alpha\, c_mB\, \hbox{Vol}_{\Sigma} \,.
\end{equation}

The factor of $\alpha$ in \eqref{E:4dresult} implies that the entanglement entropy depends on the particular Bardeen counterterm used to shift the anomaly between the gravitational sector and the gauge sector. Indeed, we expect that if the gravitational anomaly does not appear in the appropriate part of the non-conservation law for the stress tensor then the entanglement entropy would not be sensitive to it. As discussed earlier, the particular choice of $\alpha$ in a given theory depends on how the currents are coupled to dynamical gauge fields or to the metric. Much like the canonical axial-vector anomaly, if we plan on coupling $J^{\mu}$ to a dynamical gauge field then we will be forced to set $\alpha=1$.

\subsection{Six-dimensional theories}
\label{SS:sixd}
Six-dimensional theories have two types of gravitational anomalies and a mixed gauge-gravitational anomaly. We classify the Chern-Simons terms associated with these anomalies by
\begin{align}
\begin{split}
    c_m\,\bm{F}^2\, \hbox{Tr} \left(\bm{R}^2\right) & = c_m\,d\left[ (1-\alpha) \bm{A} \bm{F} \,\hbox{Tr}\left(\bm{R}^2\right) + \alpha \bm{F}^2 \hbox{Tr} \left(\bm{\Gamma}\bm{R} - \frac{1}{3} \bm{\Gamma}^3 \right)\right] = d \bm{I}_{7,m} \,,\\
    c_a\,\hbox{Tr} \left(\bm{R}^2\right)^2 & = c_a\,d \left[\hbox{Tr} \left(\bm{\Gamma}\bm{R} - \frac{1}{3} \bm{\Gamma}^3 \right) \,\hbox{Tr}\left(\bm{R}^2 \right) \right] = d \bm{I}_{7,a} \,,\\
    c_b\,\hbox{Tr} \left(\bm{R}^4\right) & = c_b\,d\, \hbox{Tr} \left[\bm{R}^3 \bm{\Gamma} - \frac{2}{5} \bm{R}^2 \bm{\Gamma}^3 - \frac{1}{5} \bm{R} \bm{\Gamma}^2 \bm{R} \bm{\Gamma} + \frac{1}{5} \bm{R} \bm{\Gamma}^5 - \frac{1}{35} \bm{\Gamma}^7 \right] = d \bm{I}_{7,b} \,.
\end{split}
\end{align}
Here
\begin{align}
	c_m = \frac{1}{768\pi^3} \sum_i \chi_i q_i^2 \,,
	\qquad
	c_a = -\frac{1}{36864 \pi^3} \sum_i \left(\chi_i - 8 t_i\right) \,,
	\qquad
	c_b = -\frac{1}{46080 \pi^3} \sum_i \left(\chi_i + 28 t_i \right) \,,
\end{align}
where $\chi_i$ and $q_i$ denote the chirality and charge of the fermion species and $t_i$ counts the number of self-dual (and anti-self-dual) two-form fields.

Using \eqref{E:deftau} we find that 
\begin{subequations}
\label{E:tau6}
\begin{align}
		\CT^{\mu}{}_{\nu} =  c_m (\CT_m)^{\mu}{}_{\nu} + c_a (\CT_a)^{\mu}{}_{\nu} + c_b (\CT_b)^{\mu}{}_{\nu} \,,
\end{align}
where 
\begin{align}\label{E:tau62}
		\begin{aligned}
			(\CT_m)^{\mu}{}_{\nu} &=  \frac{\alpha}{4} \epsilon^{\alpha\beta\gamma\delta\kappa\eta} (\partial_\alpha\Gamma^\mu{}_{\beta\nu} ) \,F_{\gamma\delta}F_{\kappa\eta} \,,\\
			(\CT_a)^{\mu}{}_{\nu} &=  \frac{1}{4} \epsilon^{\alpha\beta\gamma\delta\kappa\eta} (\partial_\alpha\Gamma^\mu{}_{\beta\nu})\, R^{\rho}{}_{\lambda\gamma\delta}R^{\lambda}{}_{\rho\kappa\eta}  \,, \\
			(\CT_b)^{\mu}{}_{\nu} &= \frac{1}{2}\, \epsilon^{\alpha\beta\gamma\delta\kappa\eta} R^{\mu_1}{}_{\mu_2 \alpha\beta} \bigg(\frac{1}{4}  R^{\mu_2}{}_{\mu_3\gamma\delta} R^{\mu_3}{}_{\mu_1\kappa\eta}
    - \frac{7}{10} R^{\mu_2}{}_{\mu_3\gamma\delta}\Gamma^{\mu_3}{}_{\kappa\mu_4}\Gamma^{\mu_4}{}_{\eta\mu_1} \\
    	&\qquad\qquad\qquad\qquad\qquad\qquad\qquad\qquad\quad +\Gamma^{\mu_2}{}_{\gamma\mu_3}\Gamma^{\mu_3}{}_{\delta\mu_4}\Gamma^{\mu_4}{}_{\kappa\mu_5}\Gamma^{\mu_5}{}_{\eta\mu_1}
    \bigg) \,.
		\end{aligned}
\end{align}
\end{subequations}

Let us consider a manifold \eqref{E:ourmanifoldC} which is of the form $\mathbb{R}^{2} \times \mathcal{N}$. Since the Riemann tensor two-form and the Christoffel connection one-form are going to be block diagonal then $(\CT_b)^{\mu}{}_{\nu} $ given in \eqref{E:tau62} will vanish.
Taking advantage of the lesson learned from the analysis of the mixed anomaly in four dimensions we consider a manifold \eqref{E:ourmanifoldC} with a constant magnetic flux through $\mathcal{N}$, i.e.,
\begin{equation}
    \int_{\mathcal{N}} \bm{F}^2
    = \Phi\,.
\end{equation}
Then, integrating $\mathcal{T}^{\mu}{}_{\nu}$ over $\mathcal{N}$ we obtain an expression similar to the two-dimensional result \eqref{E:Tau2d} but with $c_g$ replaced by
$
    \alpha c_m \Phi - 24 \pi^2 c_a \tau[\Sigma]
$,
where $\tau[\Sigma]$ is the Hirzebruch signature on the entangling surface,
\begin{equation}
    \tau[\Sigma] = -\frac{1}{24\pi^2}\int_{\mathcal{N}} \text{Tr}\left(\bm{R}^2\right)\,.
\end{equation}
Thus,
\begin{align}
	\partial_\theta S_n = 2\pi i \left(1+\frac{1}{n}\right) \left(\alpha c_m \Phi - 24 \pi^2 c_a \tau[\Sigma]\right) \,,
\end{align}
and
\begin{align}
	\partial_\theta S_A = 4\pi i \left(\alpha c_m \Phi - 24 \pi^2 c_a \tau[\Sigma]\right) \,.
\end{align}

To clarify the role of $\Phi$ and $\tau[\Sigma]$, let us consider two possible setups. If $\mathcal{N} = T^4$, $F_{23}=-F_{32} = B_1$, $F_{45}=-F_{54} = B_2$ and the remaining $F$'s vanishing, we obtain $\tau[\Sigma] =0$ and $\Phi = B_1 B_2 \hbox{Vol}_\Sigma$ with $\hbox{Vol}_{\Sigma}$ the volume of the entangling surface. Likewise, if $\bm{F}=0$ and $\mathcal{N} = K3$ then $\tau[\Sigma]=-16$ and $\Phi=0$.

It seems that in order to observe the effect of the second type of gravitational anomaly the topology of the $n$-fold cover of $\mathcal{M}$ must not reduce to a (direct) product manifold. While it is straightforward to conjure a manifold for which the last line of \eqref{E:tau62} does not vanish, its interpretation as an $n$-fold cover of a manifold $\CM$ and therefore its relation to entanglement entropy is somewhat obscure. We discuss this case further in section \ref{S:discussion}.

\subsection{Higher dimensions}
\label{SS:higherd}
We are now in a position to compute the entanglement entropy associated with an even $d$-dimensional theory. The anomaly polynomial for the theories we are considering will be characterized by a set of integers $k_i$ and $m^i_n$ and take the form of
\begin{equation}
\label{E:Iddef}
    \sum_i c_i\, \bm{F}^{k_i} \prod_{n=1} \hbox{Tr}\left(\bm{R}^{2n}\right)^{m^i_n} = d \bm{I}_{d+1} \,,
\end{equation}
where for each $i$, $2 k_i + \sum_n 4 n m^i_n = d+2$.
The analysis of the previous sections implies that for a manifold of the form $\mathbb{R}^{1,1}\times\mathcal{N}$, we will have
\begin{equation}
\label{E:Snalld}
    \partial_{\theta}S_n = 2\pi i\left(1+\frac{1}{n}\right)\, C \,,
\end{equation}
where
\begin{equation}
\label{E:CmandCg}
    C  = \sum_{i,\,m^i_1 \geq 1} \hat{c}_i \int_{\mathcal{N}} \bm{F}^{k_i}\, \hbox{Tr}(\bm{R}^2)^{m^i_1-1}\prod_{n=2} \hbox{Tr}(\bm{R}^{2n})^{m^i_n} \,,\\
\end{equation}
and
\begin{equation}
    \hat{c}_i = \begin{cases}
        c_i \,,& k_i = 0 \,,\\
        \alpha_i c_i \quad \hbox{(no sum)}\,,& k_i >0\,,
    \end{cases}
\end{equation}
where the $\alpha_i$ specifies a free parameter which determines in which sector the mixed anomaly will be manifest.
Equation \eqref{E:Snalld} leads to
\begin{equation}
    \partial_{\theta}S_A = 4\pi i\, C \,.
\end{equation}

One can single out the contribution of a particular coefficient $\hat c_j$ with $m^j_1>0$ in \eqref{E:Iddef} by an appropriate choice of the manifold $\mathcal{N}$. Since the presence of anomalies is inherently related to chiral fermions (or self-dual $p$-form fields) it would be convenient to choose an $
\mathcal{N}$ which can be endowed with a spin (or string) structure.\footnote{We thank K.\,Jensen and Y.\,Tachikawa for discussions on this point.} 
For instance, choosing a spin manifold
\begin{align}
	\CN = \mathbb{R}^{2k_j} \times  K3^{m^1_n-1} \times \prod_{n=2}  \left(\mathbb{HP}^{n}\right)^{m^j_n} \,,
\end{align}
implies that $S_n$ and $S_A$ will be proportional to $\hat c_j$ only.
Using our former result for a $K3$ surface
\begin{align}
	\int_{K3} \text{Tr} \left(\bm{R}^{2}\right) = 96 (2\pi)^2 \,,
\end{align}
and
\begin{align}
	\int_{\BH\BP^{n}} \text{Tr} \left(\bm{R}^{2n}\right) = 2(2n+2 - 4^n) (-4\pi^2)^{n} \,,
\end{align}
for quaternion projective spaces \cite{massey1962non},
we find that
\begin{equation}
\label{E:Cdef}
    C = \hat c_j \Phi_j \, \left(96 (2\pi)^2\right)^{m_1^j -1} \prod_{n=2} \left(2(2n+2 - 4^n) (-4\pi^2)^{n} \right)^{m_{n}^j} \,,
\end{equation}
with $\Phi_j \equiv \int_{\BR^{2k_j}}  \bm{F}^{k_j}$.

\section{Discussion}
\label{S:discussion}

The entanglement entropy associated with a state $|\psi\rangle$ and entangling region $\Sigma$ in a quantum field theory is defined by \eqref{EE_Def}
where $\rho_A$ is the reduced density matrix computed at a particular instant of time, $t=t_0$. If the state $|\psi\rangle$ is an eigenvalue of the Hamiltonian $H$ then its time evolution is trivial then the entanglement entropy will be independent of $t_0$. Likewise, if $|\psi\rangle$ transforms trivially under boosts, one expects that $S_A$ will be invariant under boosts as well. More precisely, we expect that two inertial observers using identical entangling surfaces and cutoff schemes will agree on the entanglement entropy $S_A$.

To make a sharper statement about the boost invariance of the entanglement entropy let us take a closer look at the comparison of the entanglement entropies between two inertial observers, call them $a$ and $b$. Let observer $b$ move at constant rapidity $\kappa$ relative to observer $a$. Both observers are interested in computing the entanglement entropy associated with an entangling surface $\Sigma$ and both observers use the same cutoff scheme. Suppose observer $a$ computes the entanglement entropy $S_A(0)$ in his own frame and the expected entanglement entropy $S_A(\kappa)$ in the frame of $b$.

In his own frame, $S_A(0)$ can be computed using the replica trick \eqref{ReplicaTrick}, that is, by computing the partition function on $\mathcal{M}_n$, the $n$-fold cover of $\mathcal{M}$ where each of the sheets of $\mathcal{M}_n$ are connected along the entangling region. According to observer $a$, $S_A(\kappa)$ can be computed by considering the partition function on $\mathcal{M}_n(\theta)$, the $n$-fold cover of $\mathcal{M}$ where each of the sheets of $\mathcal{M}_n(\theta)$ are connected along an entangling region which is rotated by the angle $\theta=-i\kappa$ relative to the original. 

As we have seen, in the presence of anomalies one finds that $S_A(\kappa)$ and $S_A(0)$ are not necessarily equal. In particular, we have seen that for (Euclidean) manifolds of the form
\begin{equation}
\label{E:lineelement}
    ds^2 = dt^2 + (dx^1)^2 + ds^2_{\mathcal{N}} \,,
\end{equation}
and an entangling surface $\Sigma= \{ \vec x \,| \,t=0,x^1=0\}$, the change in entanglement entropy due to a boost is given by
\begin{equation}
\label{E:finalresult}
    \partial_{\theta}S_A = 4\pi i \,C \,,
\end{equation}
where $C$ is given by \eqref{E:Cdef} and depends on the details of the anomaly and the manifold $\mathcal{N}$. We have argued that one may choose a manifold $\mathcal{N}$ such that $C$ will be non-zero as long as the anomaly polynomial $\bm{P}$ satisfies,
\begin{equation}
\label{E:SAcriterion}
    \left(\frac{\partial \bm{P}}{\partial \hbox{Tr}\left(\bm{R}^2\right)}\right)_{\text{Tr}(\bm{R}^{2k})} \neq 0\,,
\end{equation}
i.e., it's derivative with respect to $\hbox{Tr}\left(\bm{R}^2\right)$ with all other $\hbox{Tr}\left(\bm{R}^{2k}\right)$ held fixed is non-zero. 

It is unclear whether the entanglement entropy is susceptible to anomalies for which the right hand side of \eqref{E:SAcriterion} vanishes. From the arguments presented in section \ref{SS:sixd} we expect that for the latter type of anomalies one would need to consider more intricate entangling surfaces $\Sigma$, or manifolds $\mathcal{M}$ whose structure is different from \eqref{E:lineelement}. For instance, if we use
\begin{equation}
\label{E:alternateLE}
    ds^2 = d\rho^2 + \rho^2\left(d\tau + \omega_1(y^1) dy^2 + \omega_2(y^3) dy^4\right)^2 + \sum_{i=1}^4 (dy^i)^2 \,,
\end{equation}
as the line element for a six-dimensional manifold $\mathcal{M}_n$ with $0 \leq  \tau < 2\pi n$ then it is a straightforward (though tedious) exercise to show that at least for small $\omega_i$, $\partial_{\theta}S_A$ will receive contributions from both $c_a$ and $c_b$ defined in \eqref{E:tau6}, the latter being associated with a $\hbox{Tr}(\bm{R}^4)$ term in the anomaly polynomial. It would be interesting to be able to associate a well-defined entangling surface with \eqref{E:alternateLE} or with a variant of it.

A result similar to that in \eqref{E:Snmain} appeared in \cite{Castro:2014tta} for two-dimensional conformal field theories with a finite entangling surface. By extending the standard argument relating correlation functions of twist operators to the partition function on $\mathcal{M}_n$, $Z_n = e^{-W_n}$ \cite{Calabrese:2004eu,Cardy:2007mb}, it was argued in \cite{Castro:2014tta} that
\begin{equation}
\label{E:theirZn}
	Z_n = \frac{C_n}{z^{2h} \bar{z}^{2\bar{h}}}\,,
\end{equation}
with
\begin{equation}
	h= \frac{c_L}{24} \left(n-\frac{1}{n} \right) \,,
	\qquad
	\bar{h} = \frac{c_R}{24} \left(n-\frac{1}{n} \right)\,.
\end{equation}
Here the entangling surface consists of the points $0$ and $z$ in $\mathcal{M}_n$ and $C_n$ is a constant which does not depend on $z$. Let us consider $z=R$ and carry out an infinitesimal coordinate transformation $z \to R(1+i n \delta\theta)$. On one hand, following the general analysis of section \ref{SS:twod},\footnote{%
Equation \eqref{E:ourmethod} can be obtained by using the same manipulations as those which lead to \eqref{E:finterval}. 
Note that the coordinate transformation $(t,\,x^1) \to (z,\,\bar{z})$ with $z=x^1+it$ and $\bar{z} = x^1-it$ is a linear orientation reversing coordinate transformation so that $\epsilon^{z\bar{z}} = -1/\sqrt{g}$
relative to $\epsilon^{tx^1} = 1/\sqrt{g}$. Here $\sqrt{g}$ is the square root of the determinant of the metric which is imaginary in the $z$, $\bar{z}$ coordinate system and real in the $t$, $x^1$ coordinate system.
}
the change in the partition function due to rotations by an angle $n \delta\theta$ is given by
\begin{equation}
\label{E:ourmethod}
	\frac{\delta_{\theta} Z_n}{Z_n} = -i \delta\theta (n^2-1) \frac{c_L-c_R}{24}\,.
\end{equation}
On the other hand, we have
\begin{equation}
\label{E:twistmethod}
	\frac{ \delta_{\theta} \left(\frac{C_n}{z^{2h} \bar{z}^{2\bar{h}}}\right)}{\frac{C_n}{z^{2h} \bar{z}^{2\bar{h}}}} =  -i \delta\theta (n^2-1)\frac{c_L-c_R}{12} + \frac{\delta C_n}{C_n}\,.
\end{equation}
The validity of \eqref{E:theirZn} seems to imply that
\begin{equation}
\label{E:deltaC}
	\frac{\delta{C_n}}{C_n} = i \delta\theta (n^2-1)\frac{c_L-c_R}{24}\,.
\end{equation}
A computation of correlation functions for twist operators in the absence of a gravitational anomaly has been carried out in \cite{Lunin:2000yv} where an explicit expression for $C_n$ has been obtained. By adding the contribution of the effective action for gravitational anomalies \cite{Bardeen:1984pm,Valle:2012em,Hwang:1985uj} to the analysis of \cite{Lunin:2000yv} one should be able to verify the form \eqref{E:theirZn} and derive \eqref{E:deltaC}.

Note that if we set $\delta C_n=0$ then \eqref{E:twistmethod} will differ from \eqref{E:ourmethod} by a factor of 2. Curiously, such a factor of 2 will appear in the holographic computation of the entanglement entropy if we mistreat certain terms which are integrated by parts (see the discussion in appendix \ref{A:example}). Further, we can define a \emph{covariant} entanglement entropy by using
\begin{equation}
	\delta_{\theta} W_n^{\text{(cov)}} = i\int_{\mathcal{M}_n}  \sqrt{g}\, \chi_{\mu} \tau^{\mu} \,,
\end{equation}
where
\begin{equation}
\label{E:taucov2d}
	\tau^{\mu} = {c_g}\,\epsilon^{\mu\nu}\partial_{\nu}R \,,
\end{equation}
and $R$ is the Ricci scalar. The expression \eqref{E:taucov2d} follows from the non-conservation law for the covariant stress tensor, 
\begin{equation}
	\nabla_{\mu} T^{\text{(cov)}\,\mu\nu} = -i\tau^{\nu} \,,
\end{equation}
in the absence of an external field strength $F$. We find that 
\begin{align}
\label{E:Wncov}
	\delta_{\theta} W_n^{\text{(cov)}} = 4\pi (n^2 -1)\,i\delta\theta\, c_g \,,
\end{align}
which
is twice as large as \eqref{E:EE_Consistent2d}. Inserting \eqref{E:Wncov}  and $\delta_{\theta}W_1=0$ into \eqref{E:Renyidefinition} we obtain
\begin{equation}
	\partial_{\theta}S_n^{\text{(cov)}} = 4\pi i \left(1+\frac{1}{n}\right) c_g\,.
\end{equation}

A possible resolution of the aforementioned factor of two discrepancy between the results presented here (and the holographic result) and the conformal field theory computation presented above has been advocated in \cite{IqbalWall}. Since the origin of our coordinate system $t=x^1=0$ contains a coordinate singularity it is possible that the origin contains an extra delta function contribution. While such a contribution is, perhaps, surprising given that the geometry has been regulated we can not rule out such a possibility.
As we pointed out earlier, one can construct a coordinate system such as \eqref{E:nocs} which can be used to compute the entanglement entropy and does not posses a coordinate singularity. Such a coordinate system may a priori resolve the problems raised by \cite{IqbalWall}. However, in \cite{Ohmori:2014eia} it was argued that one would expect, on physical grounds, that  the entangling region be smeared over an ultraviolet parameter $\varepsilon$ which, in essence, serves as a boundary. It may be that such arguments, when applied to anomalies, introduce contributions to the entanglement entropy which persist even as we take the limit where the boundary vanishes.
A preliminary analysis of the modification of entanglement entropy due to boosts in the presence of boundaries has been carried out in \cite{IqbalWall} where some, but not all, contributions of boundary terms have been analyzed. We postpone a full analysis of boundary terms for future work.

Our result \eqref{E:Snmain} implies that the entanglement entropy is sensitive to gravitational and mixed gauge-gravitational anomalies. It would be interesting to understand whether the entanglement entropy is also sensitive to discrete anomalies such as parity, or to gauge anomalies (or gauge transformations in the presence of mixed anomalies). 
Indeed, following an analysis similar to the one we have used so far, one expects that under a gauge transformation $\bm{A} \to \bm{A} + d \lambda$ the generating function transform as
\begin{equation}
\label{E:gaugevariation}
    \delta_{\lambda}W_n = i\int_{\CM_n} \lambda *\bm{\mathcal{J}} \,.
\end{equation}
For a $U(1)$ polygon anomaly in $d=2m$ spacetime dimensions and $\lambda$ a constant,  \eqref{E:gaugevariation} reduces to
\begin{equation}\label{GaugeAnomalyInW}
    \delta_{\lambda}W_n = i c\, \lambda \int_{\CM_n} \bm{F}^m \,,
\end{equation}
where $c$ is the strength of the anomaly.

Since we have defined $\bm{F}$ to be the (external) field strength on $\CM_1$, it is insensitive to the $n$-sheeted cover and then $\delta_{\lambda}W \propto n$ in which case $\partial_{\lambda} S_n = 0$.
Put differently, the external gauge field $\bm{A}$ has the $\BZ_n$ replica symmetry $\bm{A}(\tau=0) = \bm{A}(\tau=2\pi) = \cdots = \bm{A}(\tau=2\pi n)$ on $\CM_n$ which yields $\int_{\CM_n} \bm{F}^m = n \int_{\CM_1} \bm{F}^m$ for an integer $n$.
Inserting the former result into  \eqref{GaugeAnomalyInW} and using  \eqref{E:Renyidefinition} results in $\partial_\lambda S_n =0$ for integer $n\ge 2$ implying $\partial_\lambda S_A =0$.

A similar argument, relying also on the behavior of curvature invariants on $\mathcal{M}_n$ \cite{Fursaev:1995ef} implies that $\partial_{\lambda} S_n = 0$ in the presence of mixed anomalies for the configurations of the form \eqref{E:lineelement}. It would be interesting to consider more involved geometries for which $\partial_{\lambda} S_n$ might pick up the anomaly. Or, perhaps $\partial_{\lambda}S_n \neq 0$ if instead of considering the entanglement between spatial regions we consider the entanglement between charged degrees of freedom.

\acknowledgments
We are grateful to T.\,Azeyanagi, N.\,Iqbal, K.\,Jensen, R.\,Loganayagam, S.\,Matsuura, G.\,S.\,Ng, K.\,Ohmori, M.\,Rangamani, S.\,Razamat, S.\,Ryu, H.\,Shimizu, Y.\,Tachikawa and A.\,Wall for valuable discussions and correspondence. We would also like to thank the organizers of ``The 9th Asian winter school on strings, particles and cosmology'' for hospitality when this project was initiated. The work of TN was supported in part by JSPS Grant-in-Aid for Young Scientists (B) No.\,15K17628. The work of AY is supported by the ISF under grant numbers 495/11, 630/14 and 1981/14, by the BSF under grant number 2014350, by the European commission FP7, under IRG 908049 and by the GIF under grant number 1156/2011.
\appendix

\section{Conventions for Wick rotating}
\label{A:conventions}
In the majority of this work we carry out computations in Euclidean signature. More precisely, the time coordinate in Lorentzian signature is analytically continued to the complex plane and computations are carried out along the imaginary time axis. In this appendix we denote quantities along the real time axis with a subscript `$L$' (Lorentzian) and quantities along the imaginary time axis with a subscript `$E$' (Euclidean). In most cases, we work exclusively in Euclidean signature and the aforementioned subscripts are omitted. In cases where there is a possible ambiguity we specify explicitly which signature metric we are working with. 

Consider a theory defined on a manifold with Lorentzian signature and time coordinate $t_L$. We may analytically continue this time coordinate to a complex one $\tau$,
\begin{equation}
	\tau = t_L - i t_E \,,
\end{equation}
where $t_L$ and $t_E$ are real. We define the Euclidean theory as a restriction of the complexified time theory to imaginary time. We denote such a restriction by
\begin{equation}
	t_L \to -i t_E \,.
\end{equation}
Given an action in Lorentzian signature
\begin{equation}
	S_L = \int  dt_L\,d^{d-1}x\,\sqrt{-g_L}\, \mathcal{L}_L \,,
\end{equation}
we analytically continue it so that
\begin{equation}
	S_L \to  i S_E \,,
\end{equation}
where
\begin{equation}
\label{E:defSE}
	S_E = - \int dt_E\,d^{d-1}x\, \sqrt{g_E}\, \mathcal{L}_E \,,
\end{equation}
and
\begin{equation}
	\mathcal{L}_E(t_E) = -\mathcal{L}_L(-i t_E) \,,
\end{equation}
arranged so that a canonical kinetic term in $\CL_E$ has opposite sign of that of $\mathcal{L}_L$.

Given a Lorentzian signature metric $g_{L\,\mu\nu} = \eta_{\mu\nu}$ and a boost with rapidity $\kappa_L$, we can extend the boost to the complexified metric via $\kappa \to \kappa_L - i \kappa_E$. The angle $\kappa_E$ (denoted by $\theta$ in the main text) will act as a rotation along the complex time coordinate and the sign is chosen so that $\kappa_E$ corresponds to a counterclockwise rotation in appropriate coordinates. For instance, we find that under boosts, $z_L=x-t$ is transformed to $\tilde{z}_L =  z_Le^{-\kappa_L}$ in Lorentzian signature while its analytic continuation $z_E=x+i t_E$ is transformed into $\tilde{z}_E = z_E e^{i\kappa_E}$. To emphasize this point let us consider the generator of boosts in Lorentzian signature 
\begin{equation}
	\chi_L^{\mu}\partial_{\mu}  = t \partial_x + x \partial_t \,.
\end{equation}
Analytically continuing to $\tau = t_L - i t_E$ and setting $t_L=0$ we find
\begin{equation}
	\chi_E^{\mu}\partial_{\mu} = -i \left(t_E \partial_x - x \partial_{t_E}\right)\,.
\end{equation}
Thus,
\begin{equation}
\label{E:boostconvention}
	\kappa_L \chi_L^{\mu}\partial_{\mu} \to  -\kappa_E \left( t_E \partial_x - x \partial_{t_E} \right)\,.
\end{equation}

A Chern-Simons action analytically continued to the imaginary time axis will become imaginary. Let us define
\begin{equation}
	\epsilon_L^{\mu_1\ldots\mu_d} = \frac{e^{\mu_1\ldots\mu_d}}{\sqrt{-g_L}} \,,
	\qquad
	\epsilon_E^{\mu_1\ldots\mu_d} = \frac{e^{\mu_1\ldots\mu_d}}{\sqrt{g_E}} \,.
\end{equation}
Given \eqref{E:defSE} we find that
\begin{equation}
	\epsilon_L^{\mu_1\ldots\mu_d} \to   i \epsilon_E^{\mu_1\ldots\mu_d}\,.
\end{equation}
Thus, given
\begin{equation}
\label{E:ChernSimonsLE}
	S_{L\,\text{CS}} = \int d^dx_L\,\sqrt{-g_L}\, \epsilon_L^{\mu_1\ldots\mu_d} \ldots \,,
\end{equation}
we have
\begin{equation}
	S_{L\,\text{CS}} \to i S_{E\,\text{CS}} \,,
\end{equation}
with
\begin{equation}
	i S_{E\,\text{CS}}=
	\int d^dx_E (-i \sqrt{g_E})\, (i \epsilon_E^{\mu_1\ldots\mu_d}) \ldots \,,
\end{equation}
resulting in
\begin{equation}
\label{E:EuclideanCS}
	S_{E\,\text{CS}} = -i \int d^dx_E\,\sqrt{g_E}\, \epsilon_E^{\mu_1\ldots\mu_d} \ldots \,.
\end{equation}

The generating function of connected correlators is given by
\begin{equation}
	Z_L[g_{L\,\mu\nu}] = \int D\phi_L \,e^{i S_L} \,.
\end{equation}
We can analytically continue this generating function to a Euclidean signature metric by extending the metric to a complex one and restricting the path integral to contributions from its Euclidean component:
\begin{equation}
	Z_L[g_{L\,\mu\nu}] \to e^{\varphi}  \int D\phi_E \,e^{-S_E} = e^{\varphi} Z_E[g_{E\,\mu\nu}] \,.
\end{equation}
The relative factor $e^\varphi$ denotes a possible phase gained by the Euclidean partition function due to the change of variables from $\phi_L$ to $\phi_E$. (For instance, the complexified time component of the gauge field is imaginary along the imaginary time axis). Such a constant phase will not affect any of our computations and we will set it to zero in what follows. Thus, we find, for example,
\begin{align}
	T^{\mu\nu}_L &= -i \frac{2}{\sqrt{-g_L}} \frac{\delta \ln Z}{\delta g_{L\,\mu\nu}} = \frac{2}{\sqrt{-g_L}} \frac{\delta W_L}{\delta g_{L\,\mu\nu}} \,,\\
	T^{\mu\nu}_E &= -\frac{2}{\sqrt{g_E}} \frac{\delta \ln Z}{\delta g_{E\,\mu\nu}} = \frac{2}{\sqrt{g_E}} \frac{\delta W_E}{\delta g_{E\,\mu\nu}}\,,
\end{align}
and hence
\begin{equation}
	W_L \to iW_E\,.
\end{equation}

\section{Anomalous Ward identities and the Chern-Simons term}
\label{A:inflow}
In this appendix we work out the Ward identities associated with the anomalous non-conservation law for the stress tensor in the presence of anomalies. We present a formal derivation of these Ward identities in section \ref{A:Ward} using the anomaly inflow mechanism. In section \ref{A:example} we rederive the Ward identity for gravitational anomalies in three dimensions in an explicit manner and in section \ref{A:holography} we connect this result to a holographic computation of entanglement entropy.

\subsection{Ward identities and anomaly inflow}
\label{A:Ward}
We have argued that in the presence of anomalies the coordinate variation of the generating function $\delta_{\chi}W$ satisfies
\begin{subequations}
\label{E:goal}
\begin{equation}
	\delta_{\chi}W = i\int_\CM \sqrt{g}\, \partial_{\mu}\chi^{\nu}\,\mathcal{T}^{\mu}{}_{\nu} \,,
\end{equation}
where
\begin{equation}\label{E:DefTauTensor}
	*\bm{\mathcal{T}}^{\mu}{}_{\nu} = \frac{\partial{\bm{I}_\text{CS}}}{\partial \bm{\Gamma}^{\nu}{}_{\mu}}\,.
\end{equation}
\end{subequations}
We will now follow \cite{Jensen:2013kka} and derive this result explicitly. As opposed to the majority of this work, the analysis will be carried out in Lorentzian signature. We then follow the conventions of appendix \ref{A:conventions} to analytically continue our results to Euclidean signature.

The anomaly inflow mechanism, discussed in \cite{Callan:1984sa}, posits that while $\delta_{\chi}W \neq 0$ for a theory with an anomaly, the variation of the covariant generating function,
\begin{equation}
	W' = W + \int_{\partial^{-1}\mathcal{M}} \bm{I}_\text{CS} \,,
\end{equation}
satisfies $\delta_{\chi}W'=0$. Here $\partial^{-1}\mathcal{M}$ is a manifold whose boundary is $\mathcal{M}$ which is where $W$ is defined and $\bm{I}_\text{CS}$ is a Chern-Simons form. Put differently, the non-coordinate invariance of $W$ is equal to the non-coordinate invariance of the Chern-Simons term,
\begin{equation}
\label{E:deltarelation}
	\delta_{\chi}W = -\delta_{\chi} \int_{\partial^{-1}\mathcal{M}} \bm{I}_\text{CS}\,.
\end{equation}

If $A^{\alpha\beta}$ is a tensor, then the change in $A^{\alpha\beta}$ under small coordinate transformations is given by its Lie derivative
\begin{equation}
	\delta_{\chi} A^{\alpha\beta} = \mathcal{L}_{\chi}A^{\alpha\beta}\,.
\end{equation}
The Christoffel connection is not a tensor and satsifies
\begin{equation}
\label{E:Gammtransformation}
	\delta_{\chi} \bm{\Gamma}^{\alpha}{}_{\beta} = \mathcal{L}_{\chi} \bm{\Gamma}^{\alpha}{}_{\beta} + d V_{\alpha}{}^{\beta} \,,
\end{equation}
where we have defined
\begin{equation}
	V_{\alpha}{}^{\beta}  = \partial_{\alpha} \chi^{\beta} \,,
\end{equation}
and
\begin{equation}
	\bm{\Gamma}^{\alpha}{}_{\beta} = \Gamma^{\alpha}{}_{\beta\gamma}dx^{\gamma}\,.
\end{equation}
Thus,
\begin{align}
\begin{split}
\label{E:deltaIV1}
	\delta_{\chi}\bm{I}_\text{CS} &= \delta_{\chi}\bm{\Gamma}^{a}{}_{b}  \frac{\partial{\bm{I}_\text{CS}}}{\partial \bm{\Gamma}^{b}{}_{a}}
		+\delta_{\chi} \bm{R}^{a}{}_{b} \frac{\partial{\bm{I}_\text{CS}}}{\partial \bm{R}^{b}{}_{a}}
		+\delta_{\chi} \bm{A} \frac{\partial{\bm{I}_\text{CS}}}{\partial \bm A}
		+\delta_{\chi} \bm{F} \frac{\partial{\bm{I}_\text{CS}}}{\partial \bm F}\,, \\
		& = d V_a{}^b  \frac{\partial{\bm{I}_\text{CS}}}{\partial \bm{\Gamma}^{b}{}_{a}} + \mathcal{L}_{\chi} \bm{I}_\text{CS}\,,
\end{split}
\end{align}
where we have written $\bm{I}_\text{CS}$ as a function of the Christoffel connection one-form, the Riemann tensor two-form $\bm{R}^{\alpha}{}_{\beta} = R^{\alpha}{}_{\beta\gamma\delta}dx^{\gamma}dx^{\delta}$ and the Abelian gauge field one-form $\bm{A}$ and its field strength $\bm{F}  = d\bm{A}$. We use roman indices for components in $\partial^{-1}\mathcal{M}$. 

In order to evaluate the Lie derivative of the Chern-Simons form we use
\begin{equation}
	\bm{I}_\text{CS} = i_\text{CS} \bm{\Omega} \,,
\end{equation}
where $\bm{\Omega}$ is the volume form on $\partial^{-1}\mathcal{M}$. Using $\mathcal{L}_{\chi}\bm{\Omega} = \bm{\Omega} \nabla_{\alpha}\chi^{\alpha}$ we have
\begin{align}
	\int \mathcal{L}_{\chi} \bm{I}_\text{CS} &= \int \partial_{\alpha} \left(\sqrt{-g}\, i_\text{CS} \chi^{\alpha} \right) d^{d+1}x =\int d\, i_{\chi} \bm{I}_\text{CS} \,,
\end{align}
where $i_{\chi}$ denotes the interior product. Thus, under the integral we have
\begin{equation}
\label{E:deltaIV2}
	\delta_{\chi}\bm{I}_\text{CS} = -V_{a}{}^{b} d \left(\frac{\partial{\bm{I}_\text{CS}}}{\partial \bm{\Gamma}^{b}{}_{a}} \right)+ d \left( V_a{}^b  \frac{\partial{\bm{I}_\text{CS}}}{\partial \bm{\Gamma}^{b}{}_{a}}  + i_{\chi} \bm{I}_\text{CS} \right)\,.
\end{equation}
Since the Chern-Simons term is invariant under coordinate transformations up to boundary terms, we conclude that $\frac{\partial{\bm{I}_\text{CS}}}{\partial \bm{\Gamma}^{b}{}_{a}}$ is closed and we are left with
\begin{equation}
\label{E:generaldeltaI}
	\delta_{\chi}\bm{I}_\text{CS} = d \left( V_a{}^b  \frac{\partial{\bm{I}_\text{CS}}}{\partial \bm{\Gamma}^{b}{}_{a}}  + i_{\chi} \bm{I}_\text{CS} \right)\,.
\end{equation}

To obtain \eqref{E:goal} we restrict ourselves to coordinate transformations for which 
\begin{subequations}
\label{E:restrictions}
\begin{align}
\label{E:restrictiona}
	\chi^{\bot} &= 0 \,,\\
\label{E:restrictionb}
	\partial_{\bot}\chi^{a} &= 0\,,
\end{align}
\end{subequations}
where $\bot$ denotes the coordinate in the bulk of $\partial^{-1}\mathcal{M}$. Having $\chi^{\bot} =0$ implies that 
\begin{equation}
	\int_{\partial^{-1}\mathcal{M}} \mathcal{L}_{\chi} \bm{I}_\text{CS} = 0	\,,
\end{equation}
and that $V_a{}^{\bot} = 0$,
while \eqref{E:restrictionb} implies that $V_\bot{}^a =  0$. Inserting these relations into \eqref{E:generaldeltaI} and using \eqref{E:deltarelation} leads us to 
\begin{align}
\label{E:finalCSvariation}
	\delta_\chi W =  - \int_{\CM} \sqrt{-g}\, \partial_{\mu}\chi^{\nu}\,\mathcal{T}^{\mu}{}_{\nu} \,,
\end{align}
with $\CT^\mu{}_\nu$ defined by \eqref{E:DefTauTensor}.

So far, we have carried out our analysis in Lorentzian signature. To go to Euclidean signature we note that each of the terms in
\begin{equation}
	W'_L = W_L + \int \bm{I}_{L\,\text{CS}} \,,
\end{equation}
is continued to Euclidean signature via
\begin{equation}
	W_L' \to i W_E \,,
	\qquad
	W_L \to i W_E \,,
	\qquad
	\int \bm{I}_{L\,\text{CS}} \to \int \bm{I}_{E\,\text{CS}}\,.
\end{equation}
(Where the relation for the Chern-Simons form follows from the second equality in \eqref{E:ChernSimonsLE}.)
Thus, the Wick rotation
\begin{equation}
	W_L + \int \bm{I}_{L\,\text{CS}} \to i W_E + \int \bm{I}_{E\,\text{CS}}\,,
\end{equation}
implies  
\begin{equation}
\label{E:deltaICSE}
	\delta_{\chi} W_E = i \delta_{\chi} \int \bm{I}_{E\,\text{CS}}\,,
\end{equation}
and leads to \eqref{E:goal} closing our argument.

The construction which we have used in obtaining \eqref{E:goal} is a purely theoretical one. The anomalous quantum field theory on $\mathcal{M}$ need not have an extension into $\partial^{-1}\mathcal{M}$. Needless to say, pion decay into two photons is not considered to be an indication for the existence of extra dimensions. For this reason, we have chosen to restrict $\chi$ according to \eqref{E:restrictions}---the theory on $\mathcal{M}$ should not be aware of the precise extension of $\chi$ from $\mathcal{M}$ into $\partial^{-1}\mathcal{M}$. Indeed, it does not seem unlikely that one may add to $W'$ boundary counterterms involving the extrinsic curvature of $\mathcal{M}$ from which one may derive \eqref{E:goal} without the use of \eqref{E:restrictions}.

\subsection{Gravitational anomalies in two dimensions}
\label{A:example}
In order to make contact with the works of \cite{Solodukhin:2005ah,Solodukhin:2005ns,Castro:2014tta,Guo:2015uqa,Azeyanagi:2015uoa} it is useful to consider an explicit example. Let us place our quantum field theory on an $n$-fold cover $\mathcal{M}_n$ with $\partial^{-1}\mathcal{M}_n$ being an asymptotically AdS space. While our intention is to relate the result of this section with computations of entanglement entropy in the context of the AdS/CFT correspondence, one can also consider the asymptotically AdS space $\partial^{-1}\mathcal{M}_n$ as a particularly useful extension of $\mathcal{M}_n$ into an extra dimension with no reference to holography.

Following the notation of \cite{Castro:2014tta} let us denote the Euclidean metric on $\partial^{-1}\mathcal{M}_{1+\epsilon}$ by
\begin{equation}
\label{E:theirls}
	ds^2 = e^{\epsilon\phi} (dt^2 + (dx^1)^2) + (g_{\bot\bot} + K_\mu x^\mu) (dx^{\bot})^2 + e^{\epsilon\phi} U_\mu dx^\mu dx^{\bot} + \ldots \,,
\end{equation}
where we have expanded the metric around $t=x^1=0$, the location of the conical deficit. Here $\mu=0,1$. The function $\phi$ serves as a regulator whose derivatives have compact support near the tip of the cone and satisfies:\footnote{The function $\phi$ satisfying \eqref{E:defbox} is given by $\phi = 2\log \rho$ where $\rho^2 \equiv t^2+ (x^1)^2$.
A choice of the regularized version of the function is $\phi = 2 \log \sqrt{\rho^2 + a^2}$ with a small parameter $a$ \cite{Dong:2013qoa}.
}
\begin{equation}
\label{E:defbox}
	\delta^{\mu\nu}\partial_\mu \partial_{\nu} \phi = 4\pi  \delta(t,x^1) \,,
\end{equation}
which is consistent with $\int_{\mathcal{M}_{1+\epsilon}} R = -4 \pi \epsilon$ with $n=1+\epsilon$  to linear order in $\epsilon$.

Following \cite{Castro:2014tta} and the convention in \eqref{E:EuclideanCS} let us evaluate the Chern-Simons term
\begin{equation}
\label{E:CS}
	\int \bm{I}_\text{CS} = c_g \int  \sqrt{g}\, i_\text{CS} = c_g \int \sqrt{g}\, \epsilon^{abc}\, \Gamma^d{}_{ae} \left(\partial_b \Gamma^e{}_{dc} + \frac{2}{3} \Gamma^e{}_{bf} \Gamma^f{}_{cd} \right)\,,
\end{equation}
on the line element \eqref{E:theirls}. 
In what follows we will often us $\epsilon^{abc} = e^{abc}/\sqrt{g}$ with $e^{abc}$ the Levi-Civita symbol satisfying $e^{t x^1 x^{\bot}} =  1$.
As a warmup exercise let us evaluate the term $\int \bm{I}_\text{CS}$ on the line element \eqref{E:theirls}. After taking into account that $\phi$ has compact support near the origin, one finds that the only contribution to terms in $\int \bm{I}_\text{CS}$ which are linear in $\epsilon$ are given by 
\begin{align}
\begin{split}
\label{E:theirSCS}
	\partial_{\epsilon} \int \bm{I}_\text{CS} \big|_{\epsilon=0} & = -\frac{1}{4} c_g \int d^2x dx^{\bot} \left[\square \phi\, e^{\mu\nu}\partial_{\mu}U_{\nu} - \partial_{\mu}\phi\, \delta^{\mu\nu}\, \partial_{\nu}( e^{\alpha\beta}\partial_{\alpha}U_{\beta})\right] \,,\\
	& = -\frac{1}{2} c_g \int d^2x dx^{\bot}\,\square \phi \,e^{\mu\nu}\partial_{\mu}U_{\nu} \,,\\
	& = -2\pi c_g \int dx^{\bot} e^{\mu\nu}\partial_{\mu}U_{\nu}\,.
\end{split}
\end{align}

We would like to see how the result \eqref{E:theirSCS} changes under a (small) $x^{\bot}$ dependent rotation in the $t$--$x^1$ plane parameterized by 
\begin{equation}
\label{E:theirchi}
	\chi^{a} = \begin{pmatrix} \delta\theta(x^{\bot})\, x^1\,, & -\delta\theta(x^{\bot})\, t\,, & 0 \end{pmatrix}\,.
\end{equation}
That is, we would like to compute $\delta_{\chi} \partial_{\epsilon} \int\bm{I}_\text{CS} \big|_{\epsilon=0}$.
Note that \eqref{E:theirchi} satisfies \eqref{E:restrictiona}  but not \eqref{E:restrictionb}. We will come back to this point later. The coordinate transformation associated with $\chi^a$ will shift the line element from \eqref{E:theirls} to 
\begin{equation}
\label{E:dstheta}
	ds^2_{\theta} = ds^2 + \delta g_{ab} dx^a dx^b \,,
\end{equation}
with
\begin{equation}
	\delta g_{ab} = \mathcal{L}_{\chi} g_{ab} \,.
\end{equation}
The rotational invariance in the $t$--$x^1$ plane allows us to write
\begin{equation}
\label{E:rotatedmetric}
	\delta g_{ab}dx^a dx^b = 	(
	\delta g_{\bot\bot} + \delta K_\mu x^\mu) (dx^{\bot})^2 + e^{\epsilon\phi} \delta U_\mu dx^\mu dx^\bot + \cdots\,.
\end{equation}
where, for example,
\begin{equation}
\label{E:deltaUval}
	\delta^{\mu\nu} \, \delta U_\nu = -\delta\theta e^{\mu\nu}U_{\nu} + 2 \delta\theta' e^{\mu\nu}\sigma_{\nu} +\delta^{\mu\nu} \,\delta\theta \partial_{\kappa}U_{\nu} e^{\kappa\rho}\sigma_{\rho} + \cdots \,.
\end{equation}

Since the rotated metric \eqref{E:rotatedmetric} has the same structure as the unrotated one \eqref{E:theirls}, we can use \eqref{E:theirSCS} directly to obtain
\begin{equation}
\label{E:shiftICS1}
	\delta_{\chi} \partial_{\epsilon} \int \bm{I}_\text{CS} \big|_{\epsilon=0}  = -\frac{1}{4} c_g \int d^2x dx^{\bot} \left[\square \phi\, e^{\mu\nu}\partial_{\mu}\delta U_{\nu} - \partial_{\mu}\phi\, \delta^{\mu\nu}\, \partial_{\nu}( e^{\alpha\beta}\partial_{\alpha} \delta U_{\beta})\right] \,.
\end{equation}
A short computation yields
\begin{equation}
	e^{\mu\nu}\partial_{\mu}\delta U_{\nu} = -4\, \delta \theta'(x^\bot) + \cdots \,,
\end{equation}
where $\cdots$ denotes terms which vanish when localized at the origin.
Therefore, the first term on the right hand side of \eqref{E:shiftICS1} will contribute to $\delta_{\chi} \int \bm{I}_\text{CS}$, but the second term will not. Thus,
\begin{align}
\begin{split}
\label{E:SCSresultinfinite}
	\delta_{\chi} \partial_{\epsilon} \int \bm{I}_\text{CS} \big|_{\epsilon=0}  =& -\frac{1}{4} c_g \int d^2x dx^{\bot}\, \square \phi\, e^{\mu\nu}\partial_{\mu}\delta U_{\nu} \,,\\
	=&4\pi c_g\, \delta\theta(x^{\bot}=0)\,,
\end{split}
\end{align}
where in the last equality we have assumed that the entangling region is a semi-infinite line, i.e., $\delta\theta(x^{\bot}=0)$ is single valued. In the case of a finite interval we obtain
\begin{equation}
\label{E:SCSresultfinite}
	\delta_{\chi} \partial_{\epsilon} \int \bm{I}_\text{CS} \big|_{\epsilon=0}  = 8 \pi c_g \delta\theta\,.
\end{equation}

The results \eqref{E:SCSresultinfinite} and \eqref{E:SCSresultfinite} may be modified since \eqref{E:theirchi} does not respect \eqref{E:restrictionb}.  As we discussed, a non-vanishing contribution to $\delta_{\chi} \partial_{\epsilon} \int \bm{I}_\text{CS}$ as a result of $\partial_{\bot}\chi^a \neq 0$ is a remnant of the construction we have been using and may be expected to cancel via boundary counterterms associated with the extrinsic curvature of $\mathcal{M}_n$, perhaps similar to the terms computed in \cite{Solodukhin:2005ah,Skenderis:2009nt,Guo:2015uqa}. Luckily, we do not have to worry about these terms: an analysis similar to the one carried out in obtaining \eqref{E:theirSCS} suggests that
\begin{equation}
	\partial_{\epsilon} \int V_{\bot}{}^{\alpha} \frac{\partial \bm{I}_\text{CS}}{\partial \bm{\Gamma}^{\alpha}{}_{\bot}} = 0 \,,
\end{equation}
implying that no new terms are generated from the violation of \eqref{E:restrictionb}. 

Using \eqref{E:deltaICSE} we find that the variation of the Euclidean generating function is given by
\begin{equation}\label{E:VariedCSaction}
	 \delta_{\chi} \partial_{\epsilon}W_{1+\epsilon}\big|_{\epsilon=0}  = i \delta_{\chi} \partial_{\epsilon} \int \bm{I}_\text{CS} \big|_{\epsilon=0} = 4 \pi i\, c_g\delta\theta\,,
\end{equation}
in precise agreement with \eqref{E:EE_Consistent2d} in the limit where $n=1+\epsilon$. For a finite interval we would have obtained
\begin{equation}\label{E:VariedCSaction_Interval}
	 \delta_{\chi} \partial_{\epsilon}W_{1+\epsilon}\big|_{\epsilon=0}  = 8 \pi i\, c_g\delta\theta\,.
\end{equation}
in agreement with our field theory expectations.

\subsection{Holography}
\label{A:holography}
We may also use \eqref{E:SCSresultfinite} to compute the change in entanglement entropy of an interval due to a boost as predicted by holography \cite{Ryu:2006bv,Lewkowycz:2013nqa,Dong:2013qoa}.
Consider a bulk action of the form
\begin{equation}
	S_\text{bulk} = S_\text{invariant} + S_\text{CS} + S_\text{boundary} \,,
\end{equation}
where $S_\text{invariant}$ denotes the fully gauge and diffeomorphism invariant part of the action, $S_\text{CS} = -i\int\bm{I}_\text{CS}$ is the contribution of the Chern-Simons term to the bulk action and $S_\text{boundary}$ denotes possible boundary terms which may contribute to the on-shell action. Following the notation of Dong \cite{Dong:2013qoa}, we decompose $S_\text{bulk}$ into a contribution coming from the tip of the regularized conical singularity and a contribution coming from outside the tip
\begin{equation}
	S_\text{bulk} = S_\text{inside}+S_\text{outside}\,.
\end{equation}
The prescription of \cite{Dong:2013qoa} for computing the entanglement entropy can be written in the form
\begin{equation}
	S_A = -\partial_{\epsilon} S_\text{inside}\,.
\end{equation}
Using \eqref{E:SCSresultfinite} we find that (for a finite interval and ignoring possible boundary)
\begin{align}
\begin{split}
	\partial_{\theta}S_A &= - \partial_{\theta}\partial_{\epsilon} S_\text{inside} \,, \\
		&= -\partial_{\theta}\partial_{\epsilon} \left(-i \int \bm{I}_\text{CS} \right) \,,\\
		&= 8 \pi i\, c_g \,,
\end{split}
\end{align}
in agreement with \eqref{E:finterval}.

\bibliographystyle{JHEP}
\bibliography{Anomaly_EE}

\providecommand{\href}[2]{#2}\begingroup\raggedright\begin{thebibliography}{10}

\bibitem{Bilal:2008qx}
A.~Bilal, {\it {Lectures on Anomalies}},
  \href{http://xxx.lanl.gov/abs/0802.0634}{{\tt arXiv:0802.0634}}.

\bibitem{Harvey:2005it}
J.~A. Harvey, {\it {Tasi 2003 Lectures on Anomalies}},
  \href{http://xxx.lanl.gov/abs/hep-th/0509097}{{\tt hep-th/0509097}}.

\bibitem{Bertlmann}
R.~A. Bertlmann, {\em Anomalies in Quantum Field Theory (The International
  Series of Monographs on Physics)}.
\newblock Clarendon Press, 2001.

\bibitem{Erdmenger:2008rm}
J.~Erdmenger, M.~Haack, M.~Kaminski, and A.~Yarom, {\it {Fluid Dynamics of
  R-Charged Black Holes}},  {\em JHEP} {\bf 01} (2009) 055,
  [\href{http://xxx.lanl.gov/abs/0809.2488}{{\tt arXiv:0809.2488}}].

\bibitem{Son:2009tf}
D.~T. Son and P.~Surowka, {\it {Hydrodynamics with Triangle Anomalies}},  {\em
  Phys. Rev. Lett.} {\bf 103} (2009) 191601,
  [\href{http://xxx.lanl.gov/abs/0906.5044}{{\tt arXiv:0906.5044}}].

\bibitem{Neiman:2010zi}
Y.~Neiman and Y.~Oz, {\it {Relativistic Hydrodynamics with General Anomalous
  Charges}},  {\em JHEP} {\bf 03} (2011) 023,
  [\href{http://xxx.lanl.gov/abs/1011.5107}{{\tt arXiv:1011.5107}}].

\bibitem{Landsteiner:2011cp}
K.~Landsteiner, E.~Megias, and F.~Pena-Benitez, {\it {Gravitational Anomaly and
  Transport}},  {\em Phys. Rev. Lett.} {\bf 107} (2011) 021601,
  [\href{http://xxx.lanl.gov/abs/1103.5006}{{\tt arXiv:1103.5006}}].

\bibitem{Landsteiner:2011iq}
K.~Landsteiner, E.~Megias, L.~Melgar, and F.~Pena-Benitez, {\it {Holographic
  Gravitational Anomaly and Chiral Vortical Effect}},
  \href{http://xxx.lanl.gov/abs/1107.0368}{{\tt arXiv:1107.0368}}.

\bibitem{Jensen:2012kj}
K.~Jensen, R.~Loganayagam, and A.~Yarom, {\it {Thermodynamics, Gravitational
  Anomalies and Cones}},  {\em JHEP} {\bf 1302} (2013) 088,
  [\href{http://xxx.lanl.gov/abs/1207.5824}{{\tt arXiv:1207.5824}}].

\bibitem{Jensen:2013kka}
K.~Jensen, R.~Loganayagam, and A.~Yarom, {\it {Anomaly Inflow and Thermal
  Equilibrium}},  {\em JHEP} {\bf 1405} (2014) 134,
  [\href{http://xxx.lanl.gov/abs/1310.7024}{{\tt arXiv:1310.7024}}].

\bibitem{Jensen:2013rga}
K.~Jensen, R.~Loganayagam, and A.~Yarom, {\it {Chern-Simons Terms from Thermal
  Circles and Anomalies}},  {\em JHEP} {\bf 1405} (2014) 110,
  [\href{http://xxx.lanl.gov/abs/1311.2935}{{\tt arXiv:1311.2935}}].

\bibitem{Kharzeev:2007jp}
D.~E. Kharzeev, L.~D. McLerran, and H.~J. Warringa, {\it {The Effects of
  Topological Charge Change in Heavy Ion Collisions: `Event by Event P and CP
  Violation'}},  {\em Nucl. Phys.} {\bf A803} (2008) 227--253,
  [\href{http://xxx.lanl.gov/abs/0711.0950}{{\tt arXiv:0711.0950}}].

\bibitem{Kharzeev:2013ffa}
D.~E. Kharzeev, {\it {The Chiral Magnetic Effect and Anomaly-Induced
  Transport}},  {\em Prog.Part.Nucl.Phys.} {\bf 75} (2014) 133--151,
  [\href{http://xxx.lanl.gov/abs/1312.3348}{{\tt arXiv:1312.3348}}].

\bibitem{Kharzeev:2015kna}
D.~E. Kharzeev, {\it {Topology, Magnetic Field, and Strongly Interacting
  Matter}},  {\em Ann.Rev.Nucl.Part.Sci.} {\bf 65} (2015) 0000,
  [\href{http://xxx.lanl.gov/abs/1501.0133}{{\tt arXiv:1501.0133}}].

\bibitem{Son:2012bg}
D.~Son and B.~Spivak, {\it {Chiral Anomaly and Classical Negative
  Magnetoresistance of Weyl Metals}},  {\em Phys.Rev.} {\bf B88} (2013) 104412,
  [\href{http://xxx.lanl.gov/abs/1206.1627}{{\tt arXiv:1206.1627}}].

\bibitem{Landsteiner:2013sja}
K.~Landsteiner, {\it {Anomalous Transport of Weyl Fermions in Weyl
  Semimetals}},  {\em Phys.Rev.} {\bf B89} (2014), no.~7 075124,
  [\href{http://xxx.lanl.gov/abs/1306.4932}{{\tt arXiv:1306.4932}}].

\bibitem{2014arXiv1412.6543L}
Q.~{Li}, D.~E. {Kharzeev}, C.~{Zhang}, Y.~{Huang}, I.~{Pletikosic}, A.~V.
  {Fedorov}, R.~D. {Zhong}, J.~A. {Schneeloch}, G.~D. {Gu}, and T.~{Valla},
  {\it {Observation of the chiral magnetic effect in ZrTe5}},  {\em ArXiv
  e-prints} (Dec., 2014) [\href{http://xxx.lanl.gov/abs/1412.6543}{{\tt
  arXiv:1412.6543}}].

\bibitem{Zhang:2015gwa}
C.~Zhang, S.-Y. Xu, I.~Belopolski, Z.~Yuan, Z.~Lin, {\em et.~al.}, {\it
  {Observation of the Adler-Bell-Jackiw Chiral Anomaly in a Weyl Semimetal}},
  \href{http://xxx.lanl.gov/abs/1503.0263}{{\tt arXiv:1503.0263}}.

\bibitem{Ohnishi:2014uea}
A.~Ohnishi and N.~Yamamoto, {\it {Magnetars and the Chiral Plasma
  Instabilities}},  \href{http://xxx.lanl.gov/abs/1402.4760}{{\tt
  arXiv:1402.4760}}.

\bibitem{Kaminski:2014jda}
M.~Kaminski, C.~F. Uhlemann, M.~Bleicher, and J.~Schaffner-Bielich, {\it
  {Anomalous Hydrodynamics Kicks Neutron Stars}},
  \href{http://xxx.lanl.gov/abs/1410.3833}{{\tt arXiv:1410.3833}}.

\bibitem{Shaverin:2014xya}
E.~Shaverin and A.~Yarom, {\it {An Anomalous Propulsion Mechanism}},
  \href{http://xxx.lanl.gov/abs/1411.5581}{{\tt arXiv:1411.5581}}.

\bibitem{Wall:2011kb}
A.~C. Wall, {\it {Testing the Generalized Second Law in 1+1 Dimensional
  Conformal Vacua: an Argument for the Causal Horizon}},  {\em Phys. Rev.} {\bf
  D85} (2012) 024015, [\href{http://xxx.lanl.gov/abs/1105.3520}{{\tt
  arXiv:1105.3520}}].

\bibitem{Castro:2014tta}
A.~Castro, S.~Detournay, N.~Iqbal, and E.~Perlmutter, {\it {Holographic
  Entanglement Entropy and Gravitational Anomalies}},  {\em JHEP} {\bf 1407}
  (2014) 114, [\href{http://xxx.lanl.gov/abs/1405.2792}{{\tt
  arXiv:1405.2792}}].

\bibitem{Guo:2015uqa}
W.-z. Guo and R.-x. Miao, {\it {Entropy for gravitational Chern-Simons terms by
  squashed cone method}},  \href{http://xxx.lanl.gov/abs/1506.0839}{{\tt
  arXiv:1506.0839}}.

\bibitem{Azeyanagi:2015uoa}
T.~Azeyanagi, R.~Loganayagam, and G.~S. Ng, {\it {Holographic Entanglement for
  Chern-Simons Terms}},  \href{http://xxx.lanl.gov/abs/1507.0229}{{\tt
  arXiv:1507.0229}}.

\bibitem{Calabrese:2004eu}
P.~Calabrese and J.~L. Cardy, {\it {Entanglement Entropy and Quantum Field
  Theory}},  {\em J. Stat. Mech.} {\bf 0406} (2004) P06002,
  [\href{http://xxx.lanl.gov/abs/hep-th/0405152}{{\tt hep-th/0405152}}].

\bibitem{Lewkowycz:2013nqa}
A.~Lewkowycz and J.~Maldacena, {\it {Generalized Gravitational Entropy}},  {\em
  JHEP} {\bf 1308} (2013) 090, [\href{http://xxx.lanl.gov/abs/1304.4926}{{\tt
  arXiv:1304.4926}}].

\bibitem{Belin:2013dva}
A.~Belin, A.~Maloney, and S.~Matsuura, {\it {Holographic Phases of Renyi
  Entropies}},  {\em JHEP} {\bf 12} (2013) 050,
  [\href{http://xxx.lanl.gov/abs/1306.2640}{{\tt arXiv:1306.2640}}].

\bibitem{Camps:2014voa}
J.~Camps and W.~R. Kelly, {\it {Generalized gravitational entropy without
  replica symmetry}},  {\em JHEP} {\bf 03} (2015) 061,
  [\href{http://xxx.lanl.gov/abs/1412.4093}{{\tt arXiv:1412.4093}}].

\bibitem{Ryu:2006bv}
S.~Ryu and T.~Takayanagi, {\it {Holographic Derivation of Entanglement Entropy
  from AdS/CFT}},  {\em Phys.Rev.Lett.} {\bf 96} (2006) 181602,
  [\href{http://xxx.lanl.gov/abs/hep-th/0603001}{{\tt hep-th/0603001}}].

\bibitem{Ryu:2006ef}
S.~Ryu and T.~Takayanagi, {\it {Aspects of Holographic Entanglement Entropy}},
  {\em JHEP} {\bf 0608} (2006) 045,
  [\href{http://xxx.lanl.gov/abs/hep-th/0605073}{{\tt hep-th/0605073}}].

\bibitem{PSP:2075184}
M.~F. Atiyah, V.~K. Patodi, and I.~M. Singer, {\it Spectral asymmetry and
  riemannian geometry. i},  {\em Mathematical Proceedings of the Cambridge
  Philosophical Society} {\bf 77} (1, 1975) 43--69.

\bibitem{PSP:2075856}
M.~F. Atiyah, V.~K. Patodi, and I.~M. Singer, {\it Spectral asymmetry and
  riemannian geometry. ii},  {\em Mathematical Proceedings of the Cambridge
  Philosophical Society} {\bf 78} (11, 1975) 405--432.

\bibitem{PSP:2128256}
M.~F. Atiyah, V.~K. Patodi, and I.~M. Singer, {\it Spectral asymmetry and
  riemannian geometry. iii},  {\em Mathematical Proceedings of the Cambridge
  Philosophical Society} {\bf 79} (1, 1976) 71--99.

\bibitem{IqbalWall}
N.~Iqbal and A.~Wall, ``Anomalies of the entanglement entropy in chiral
  theories.'' to appear.

\bibitem{Holzhey:1994we}
C.~Holzhey, F.~Larsen, and F.~Wilczek, {\it {Geometric and Renormalized Entropy
  in Conformal Field Theory}},  {\em Nucl.Phys.} {\bf B424} (1994) 443--467,
  [\href{http://xxx.lanl.gov/abs/hep-th/9403108}{{\tt hep-th/9403108}}].

\bibitem{Solodukhin:2008dh}
S.~N. Solodukhin, {\it {Entanglement entropy, conformal invariance and
  extrinsic geometry}},  {\em Phys. Lett.} {\bf B665} (2008) 305--309,
  [\href{http://xxx.lanl.gov/abs/0802.3117}{{\tt arXiv:0802.3117}}].

\bibitem{Myers:2010tj}
R.~C. Myers and A.~Sinha, {\it {Holographic C-Theorems in Arbitrary
  Dimensions}},  {\em JHEP} {\bf 1101} (2011) 125,
  [\href{http://xxx.lanl.gov/abs/1011.5819}{{\tt arXiv:1011.5819}}].

\bibitem{Casini:2011kv}
H.~Casini, M.~Huerta, and R.~C. Myers, {\it {Towards a Derivation of
  Holographic Entanglement Entropy}},  {\em JHEP} {\bf 1105} (2011) 036,
  [\href{http://xxx.lanl.gov/abs/1102.0440}{{\tt arXiv:1102.0440}}].

\bibitem{Fursaev:2013fta}
D.~V. Fursaev, A.~Patrushev, and S.~N. Solodukhin, {\it {Distributional
  Geometry of Squashed Cones}},  {\em Phys. Rev.} {\bf D88} (2013), no.~4
  044054, [\href{http://xxx.lanl.gov/abs/1306.4000}{{\tt arXiv:1306.4000}}].

\bibitem{Bardeen:1984pm}
W.~A. Bardeen and B.~Zumino, {\it {Consistent and Covariant Anomalies in Gauge
  and Gravitational Theories}},  {\em Nucl. Phys.} {\bf B244} (1984) 421.

\bibitem{Jensen:2013vta}
K.~Jensen, P.~Kovtun, and A.~Ritz, {\it {Chiral Conductivities and Effective
  Field Theory}},  {\em JHEP} {\bf 1310} (2013) 186,
  [\href{http://xxx.lanl.gov/abs/1307.3234}{{\tt arXiv:1307.3234}}].

\bibitem{Fursaev:1995ef}
D.~V. Fursaev and S.~N. Solodukhin, {\it {On the Description of the Riemannian
  Geometry in the Presence of Conical Defects}},  {\em Phys. Rev.} {\bf D52}
  (1995) 2133--2143, [\href{http://xxx.lanl.gov/abs/hep-th/9501127}{{\tt
  hep-th/9501127}}].

\bibitem{massey1962non}
W.~Massey {\em et.~al.}, {\it Non-existence of almost-complex structures on
  quaternionic projective spaces},  {\em Pacific J. Math} {\bf 12} (1962)
  1379--1384.

\bibitem{Cardy:2007mb}
J.~L. Cardy, O.~A. Castro-Alvaredo, and B.~Doyon, {\it {Form factors of
  branch-point twist fields in quantum integrable models and entanglement
  entropy}},  {\em J. Statist. Phys.} {\bf 130} (2008) 129--168,
  [\href{http://xxx.lanl.gov/abs/0706.3384}{{\tt arXiv:0706.3384}}].

\bibitem{Lunin:2000yv}
O.~Lunin and S.~D. Mathur, {\it {Correlation functions for $M^N/ S_N$
  orbifolds}},  {\em Commun. Math. Phys.} {\bf 219} (2001) 399--442,
  [\href{http://xxx.lanl.gov/abs/hep-th/0006196}{{\tt hep-th/0006196}}].

\bibitem{Valle:2012em}
M.~Valle, {\it {Hydrodynamics in 1+1 dimensions with gravitational anomalies}},
   {\em JHEP} {\bf 08} (2012) 113,
  [\href{http://xxx.lanl.gov/abs/1206.1538}{{\tt arXiv:1206.1538}}].

\bibitem{Hwang:1985uj}
D.~S. Hwang, {\it {Gauge and Gravitational Anomalies in Two-dimensions}},  {\em
  Phys. Rev.} {\bf D35} (1987) 1268.

\bibitem{Ohmori:2014eia}
K.~Ohmori and Y.~Tachikawa, {\it {Physics at the entangling surface}},  {\em J.
  Stat. Mech.} {\bf 1504} (2015), no.~4 P04010,
  [\href{http://xxx.lanl.gov/abs/1406.4167}{{\tt arXiv:1406.4167}}].

\bibitem{Callan:1984sa}
J.~Callan, Curtis~G. and J.~A. Harvey, {\it {Anomalies and Fermion Zero Modes
  on Strings and Domain Walls}},  {\em Nucl.Phys.} {\bf B250} (1985) 427.

\bibitem{Solodukhin:2005ah}
S.~N. Solodukhin, {\it {Holography with gravitational Chern-Simons}},  {\em
  Phys. Rev.} {\bf D74} (2006) 024015,
  [\href{http://xxx.lanl.gov/abs/hep-th/0509148}{{\tt hep-th/0509148}}].

\bibitem{Solodukhin:2005ns}
S.~N. Solodukhin, {\it {Holographic description of gravitational anomalies}},
  {\em JHEP} {\bf 07} (2006) 003,
  [\href{http://xxx.lanl.gov/abs/hep-th/0512216}{{\tt hep-th/0512216}}].

\bibitem{Dong:2013qoa}
X.~Dong, {\it {Holographic Entanglement Entropy for General Higher Derivative
  Gravity}},  {\em JHEP} {\bf 01} (2014) 044,
  [\href{http://xxx.lanl.gov/abs/1310.5713}{{\tt arXiv:1310.5713}}].

\bibitem{Skenderis:2009nt}
K.~Skenderis, M.~Taylor, and B.~C. van Rees, {\it {Topologically Massive
  Gravity and the AdS/CFT Correspondence}},  {\em JHEP} {\bf 09} (2009) 045,
  [\href{http://xxx.lanl.gov/abs/0906.4926}{{\tt arXiv:0906.4926}}].

\end{thebibliography}\endgroup

\end{document}